\documentclass{aa}  

\usepackage{ulem}
\usepackage{graphicx}
\usepackage{longtable}
\usepackage{txfonts}
\begin{document}

   \title{Exploring the properties of low-frequency radio emission and magnetic fields in a sample of compact                   galaxy groups using the LOFAR Two-Metre Sky Survey (LoTSS)}

   \subtitle{}

   \author{B. Nikiel-Wroczy\'nski       \inst{1}
          \and A. Berger                        \inst{2}
          \and N. Herrera Ruiz          \inst{2}
          \and D.~J. Bomans                     \inst{2}
          \and S. Blex                          \inst{2}
          \and C. Horellou                      \inst{3}
          \and R. Paladino                      \inst{4}
          \and A. Becker                        \inst{2}
          \and A. Miskolczi             \inst{2}
          \and R. Beck                          \inst{5}
          \and K. Chy\.zy                       \inst{1}
          \and R.-J. Dettmar            \inst{2}   
          \and G. Heald                         \inst{6}
          \and V. Heesen                        \inst{7}
          \and M. Jamrozy                       \inst{1}
                  \and T. W. Shimwell           \inst{8}
                  \and C. Tasse                         \inst{9}
          }

   \institute{
                        Astronomical Observatory, Jagiellonian University, ul. Orla 171, 30-244 Krak\'ow, Poland
         \and
             Fakult\"at f\"ur Physik und Astronomie, Astronomisches Institut, Ruhr-Universität Bochum, 44780 Bochum, Germany
                \and 
                         Chalmers University of Technology, Dept of Space, Earth and Environment, Onsala Space Observatory,
                         439 92 Onsala, Sweden
        \and 
                 INAF/Istituto di Radioastronomia, via Gobetti 101, 40129 Bologna, Italy
         \and
                         Max-Planck-Institute f\"ur Radioastronomie, Auf dem H\"ugel 69, 53121 Bonn, Germany
         \and 
                 CSIRO Astronomy and Space Science, PO Box 1130, Bentley WA 6102, Australia
         \and
                         University of Hamburg, Hamburger Sternwarte, Gojenbergsweg 112, 21029 Hamburg, Germany
        \and 
                 ASTRON, the Netherlands Institute for Radio Astronomy, Postbus 2, 7990 AA Dwingeloo, The                                        Netherlands 
        \and 
                GEPI, Observatoire de Paris, CNRS, Universit\'e Paris Diderot, 5 place Jules Janssen, 92190 Meudon,                France
             }
             
   \date{Received XXX; accepted XXX}

 \abstract{ 
We use the LOFAR Two-metre Sky Survey (LoTSS) Data Release I to identify the groups of galaxies (and individual galaxies) from the Hickson Compact Groups and Magnitude Limited Compact Groups samples that emit at the frequency of 150\,MHz, characterise their radio emission (extended or limited to the galaxies), and compare new results to earlier observations and theoretical predictions. 
The detection of 73 systems (and 7 more -- probably) out of 120, of which as many as 17 show the presence of extended radio structures, confirms the previous hypothesis of the common character of the magnetic field inside galaxy groups and its detectability. In order to investigate the future potential of low-frequency radio studies of galaxy groups, we also present a more detailed insight into four radio-emitting systems, for which the strength of the magnetic field inside their intergalactic medium (IGM) is calculated. The estimated values are comparable to that found inside star-forming galaxies, suggesting a dynamical and evolutionary importance of the magnetic field in galaxy groups.
} 
\titlerunning{LoTSS of GGroups}
   \maketitle
%

\section{Introduction}
\label{sec_intro}

Galaxy groups -- systems where only a few galaxies are gravitationally bound together, contrary to the more abundant clusters -- are believed to be so common in the Universe that it is likely that most of the existing galaxies reside inside them \citep{mulchaey00}. While originally expected to express physical conditions similar to the central regions of rich clusters \citep{hickson82}, it was later found that there are more differences between these two classes than common traits. Groups are not as depleted of neutral hydrogen as clusters are and form stars more vigorously (e.g. \citealt{huchtmeier97}). Furthermore, studies by \citet{lmw10, lmw12, lmw13} showed that infrared colour-colour plots for galaxies in groups, in clusters, and in the field are different: there is a visible deficiency of "normal" star-forming galaxies inside the groups, while both "dead" and starbursting galaxies are common; this behaviour is seen neither in field galaxies, nor in galaxies located in the centres of clusters. Recently, \citet{paul17} have presented results of numerical simulations of a sample of low- and high-mass galaxy systems (thus, groups and clusters) that revealed that none of the hydrostatic laws can be scaled from clusters down to the groups owing to a breakdown in the relations that describe these laws. The conditions at which this breakdown occurs are mass below $ 8\times 10^{13} M_{\odot}$, temperature lower than 1\,keV, and radius not larger than 1\,Mpc, which clearly separates groups from clusters. Also, the storage of non-thermal energy and turbulence in groups seem to be larger than previously thought. Last but not least, it is suggestive that groups are, in general, non-virialised systems. As a result, dynamics and evolution of any galaxy group -- no matter if poor or abundant -- cannot be extrapolated using the relations derived for clusters, and is much more dependent on non-thermal than thermal processes. This draws attention to non-thermal radio emission and the associated magnetic fields.\\

Contrary to the case of field galaxies, or galaxy clusters, properties of radio emission and magnetic fields of galaxy groups are still scarcely studied. An early study of the radio emission from a subset of the Hickson Compact Groups (HCG; \citealt{hickson82}) by \citet{menon85} revealed an extended radio structure only in one system, HCG\,60,  that was mainly due to the presence of a radio galaxy. A handful of groups have been studied in detail (e.g. \citealt{kantharia05,bnw13B}). Some of these works only considered the active galactic nuclei (AGN) role and feedback \citep{giacintucci11}, not the existence of magnetic fields inside. However, it was possible to state the preliminary conclusion that magnetic fields, wherever revealed, seem to possess enough energy to be non-negligible agents in the dynamics and evolution of their host systems \citep{bnw17}. Studies on the radio emission of groups and their member galaxies are also of high importance because of the uniqueness of the environment. As stated by \citet{hickson92}, a plenitude of phenomena that are usually not seen in field galaxies, such as morphological distortions or starburst activity, can be found in these objects. Amplification of magnetic field in interacting galaxies, with an increase to the point of nuclear coalescence, has been shown for galaxy pairs \citep{drzazga11} but not for groups, where the interaction history is likely to be more complicated. Since it is not possible to relate the results acquired when studying clusters, gathering a larger sample of testbed groups is of high interest especially at lower radio frequencies, where the thermal contribution is expected to be negligible and the low-energy relativistic electron population can still be discovered. Old, high-energy particles do not manifest themselves at higher frequencies because of the severe energy losses already encountered.\\

Construction of new instruments orientated towards low-frequency radio observing facilitated a number of successful studies. Extended haloes of galaxies  have been discovered at 150\,MHz (\citealt{mulcahy14}, \citealt{mulcahy18}, \citealt{heesen18}) along with large-scale, radio continuum structures in clusters, where previously unknown, steep-spectrum relics have been found \citep{vanweeren16, hoang17}. Whereas there has been some success in finding groups that host radio-emitting structures with LOFAR \citep{LOFAR}, such as the discovery of a giant radio galaxy embedded in a galaxy group \citep{clarke17} or a study of a sample of Fanaroff-Riley type II (\ion{FR}{II}; \citealt{fr74}) radio galaxies in groups (by \citealt{croston17}), no approach to galaxy groups as the main field of interest was made. As structures that are expected to be found in the intra-group space share similarities with intergalactic filaments, tails, and envelopes -- structures that LOFAR has already revealed in clusters and single objects -- it is clear that surveying radio-emitting galaxy groups at metrewaves with this instrument should be possible.\\

This work is a first attempt to investigate a larger sample of galaxy groups at the metrewaves and to provide a more detailed insight into the radio emission of a small subsample of radio-emitting groups, by exploiting newly available data from the \textbf{Lo}far \textbf{T}wo-metre \textbf{S}ky \textbf{S}urvey (LoTSS; \citealt{shimwell17}). With the acquired data, we aim to show that it is possible and worthwhile to conduct deep, high fidelity radio observations of galaxy groups even at very low frequencies. The paper is organised as follows: in Sect.~\ref{sec_data} the data and analysis methods are described, in Sect.~\ref{sec_results} radio emission of groups is classified on the basis of its character (intra- or intergalactic etc.), and in Sect.~\ref{sec_discussion} a more detailed insight into several examples of radio-emitting galaxy groups is provided and their total magnetic field is discussed. Analysis of the number and type of detections is also provided there. The last section, Sect.~\ref{sec_conclusions}, serves as a recapitulation of the main findings of our paper.

\section{The data}
\label{sec_data}

The low-frequency data used for this project were acquired from the LoTSS and have undergone standard calibration and reduction procedure (as outlined in \citealt{shimwell17}). To avoid the risk of blurring the emission from multiple entities into one, we chose  data sets with higher, 6 arcsec resolution. The radio images, together with information about the r.m.s. noise level ($\sigma$) of the particular LoTSS-HETDEX fields, were processed by a script that extracted smaller cutouts centred on each of the groups and superimposed radio contours on a background optical map -- an RGB mosaic composed of the \textit{gri}-band images of the Sloan Digital Sky Survey Data Release 14 (SDSS DR14; \citealt{sdss14}) data. Four particular objects were chosen for a more detailed study. In their case, lower (20 arcsecs) resolution data from the LoTSS were also analysed using \textsc{blobcat} \citep{blobcat} software. The total flux density of these galaxy groups was calculated on the basis of manual inspection of the masks that \textsc{blobcat} generated. If multiple structures were present, for example galaxies not connected by a common envelope, the final flux density measurement was the sum of the flux density of the individual emitters. Overlays on the optical maps were also made.\\
To allow comparison of low- and high-frequency structures for the sample four objects and to calculate the integrated spectral index and magnetic field strength, we used data from two 1400\,MHz radio continuum surveys: the \textbf{N}RAO \textbf{V}LA \textbf{S}ky \textbf{S}urvey (NVSS; \citealt{nvss}), and the \textbf{F}aint \textbf{I}mages of \textbf{R}adio \textbf{S}ky at \textbf{T}wenty centimeters (FIRST; \citealt{FIRST}). For consistency, the flux density measurements were carried out in the same manner as on LoTSS maps, with a single exception that fixed $\sigma$ levels of 0.45 mJy/beam for the NVSS and 0.15 mJy/beam for the FIRST data were chosen.\\

\section{Results}
\label{sec_results}

\subsection{Radio emission from the group sample}

Our sample of galaxy groups consists of "multiple" galaxies in the area roughly covered by the first LoTSS data release from the HCG \citep[][]{hickson92} and Magnitude-Limited Compact Groups \citep[MLCG;][]{mlcg} catalogues. The first of these is the "classic" galaxy group catalogue, which, even though not all of the member objects are genuine groups \citep{hickson92}, is still regarded as a complete sample (e.g.  \citealt{diazgimenez10}). The second sample uses the 'friend-of-a-friend' algorithm coupled with the redshift information to distinguish real groups from apparent groups. Both lists contain objects that are possibly parts of larger structures, such as cluster cores, regions of enhanced galaxy density in the outskirts of clusters, or members of larger, loose groups. As the differences between groups and clusters are still under debate, we decided to keep them in the sample to allow future comparison of these two subsamples ("isolated" groups \textbf{(I)}, and those "contained" within larger structures \textbf{(C)}). The total number of systems from both catalogues in the aforementioned area that were taken into account is 120. For all objects, we measured the total flux density at 150\,MHz, identified the number of emitting galaxies, and described the type of the radio emission associated with particular systems (see Table~\ref{tab:cat_groups}). Two types of radio emission have been introduced: \textbf{G} denotes systems in which the emission only comes from galaxies and \textbf{E} those hosting an extended structure covering at least some of the members. In some cases, 
the flux measurements were compromised by the presence of artefacts (e.g. caused by a bright nearby source) or the detection was marginal. These objects were marked with an single/double asterisk sign, respectively, and the exact flux measurements are replaced with question marks. Analysis of the radio emission reveals that 73 of the studied objects (19 embedded, 57 isolated) are hosts of radio emission and an additional 7 are probable detections. 
Out of the aforementioned 73 systems, 17 (6 embedded, 11 isolated) show a presence of extended structures from small, bridge-like structures, to large envelopes encompassing whole systems (e.g.\ MLCG\,1374). Therefore, the percentage of radio-emitting groups in the LoTSS DR1 sample is equal to 61--67\%; approximately 14\% of all systems show extended, radio continuum structures. This is much more than what was found by \citet{menon85}, who were only able to detect a radio envelope around HCG\,60 (out of 88 investigated systems). Theoretical predictions by \citet{paul18} have suggested that between 3 and 21\% of galaxy groups should be visible in the radio continuum, depending on the telescope set-up and detections limits. A 14\% success rate for the LoTSS seems to be in a good agreement with this conclusion. However, one should be aware of the fact that some of the detected envelopes might be created because of the inclusion of galactic/AGN emission within the beam area; analysis at other radio frequencies is necessary to test if this is a significant effect.\\

Table~\ref{tab:cat_gals} contains a separate list of measurements for those member galaxies that were found to be radio emitting. The selected sample of systems contains 419 galaxies. Out of these, 110 are certain detections; 9 seem to be emitting, but the contribution of each of these 9 galaxies could not be separated from a larger structure; 17 are probable emitters (detection uncertain owing to the presence of artefacts and a low signal-to-noise ratio); and 283 are not emitting at all. This gives an overall fraction of radio-emitting galaxies in groups of 26--32\% (0.92--1.13 emitting galaxy per group). If taking into consideration only the radio-emitting groups, these numbers are 41--51\% and 1.38--1.79, respectively. This confirms that a significant number of galaxies in groups are indeed radio emitters.\\

\subsection{Close-up view of four sample systems}

We selected a few systems of particular interest for a more detailed analysis. The general guidelines for choosing these systems were as follows:
\begin{itemize}
\item Confident detection of an extended radio structure at 150\,MHz
\item Lack of bright radio sources that may cause strong artefacts
\item Half of the sample should come from the isolated subsample, the other half from the contained subsample\item Velocity dispersion between the group members should be low enough to undoubtedly identify them as bound together
\end{itemize}

The selected groups are HCG\,60, MLCG\,24, MLCG\,41, and MLCG\,1374. All these systems exhibit intergalactic radio continuum emission and/or clear signs of collision in the radio regime, suggesting the existence of intergalactic, magnetised structures; have a well-known set of members; and their radio emission is not affected by imaging artefacts. The first two are possible members of larger associations; the other two are believed to be isolated.

\subsubsection{Contained systems with possible AGN presence: HCG\,60 and MLCG\,24}

\begin{figure*} 
\centering
        \includegraphics[width=0.49\textwidth]{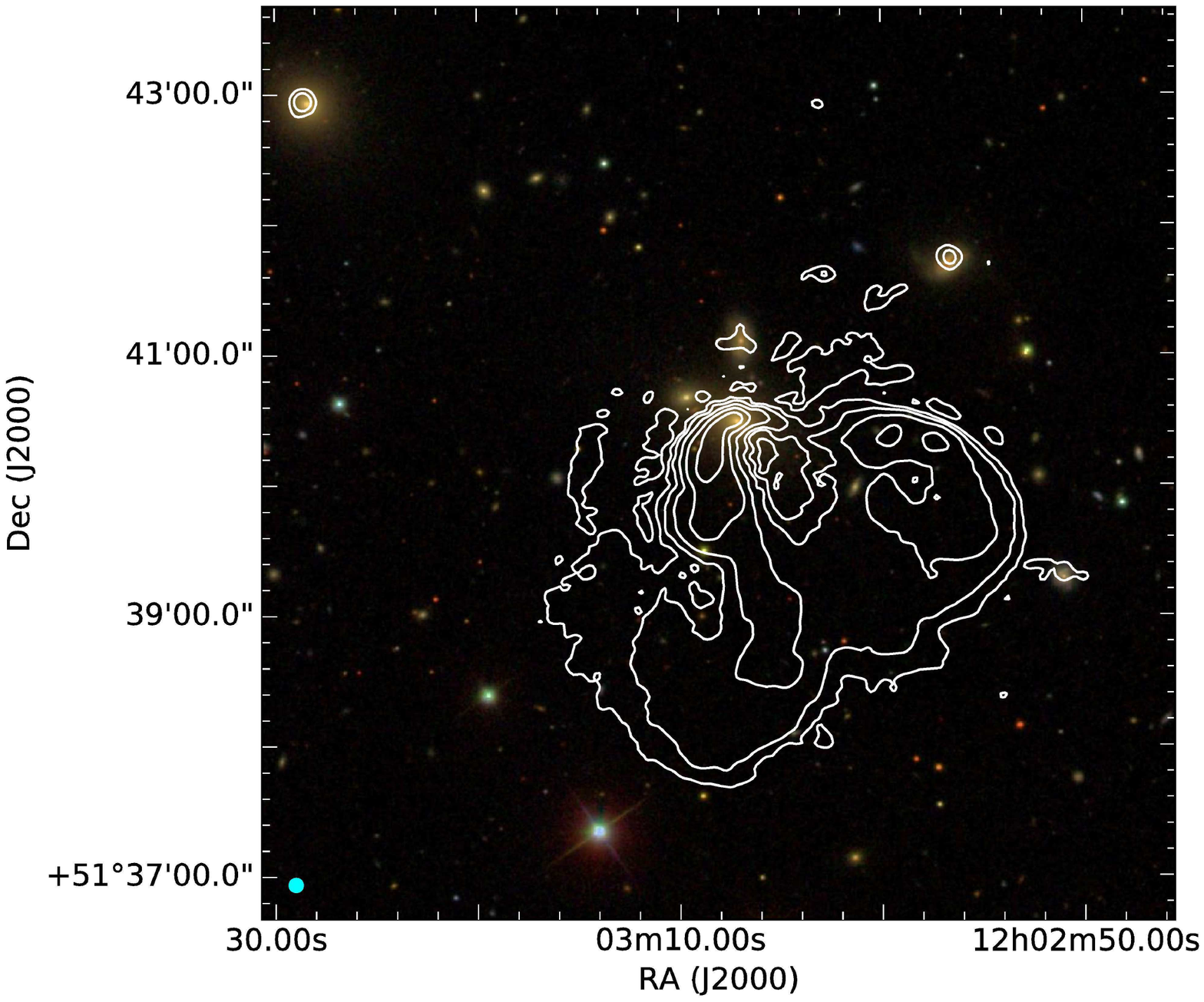}
    \includegraphics[width=0.49\textwidth]{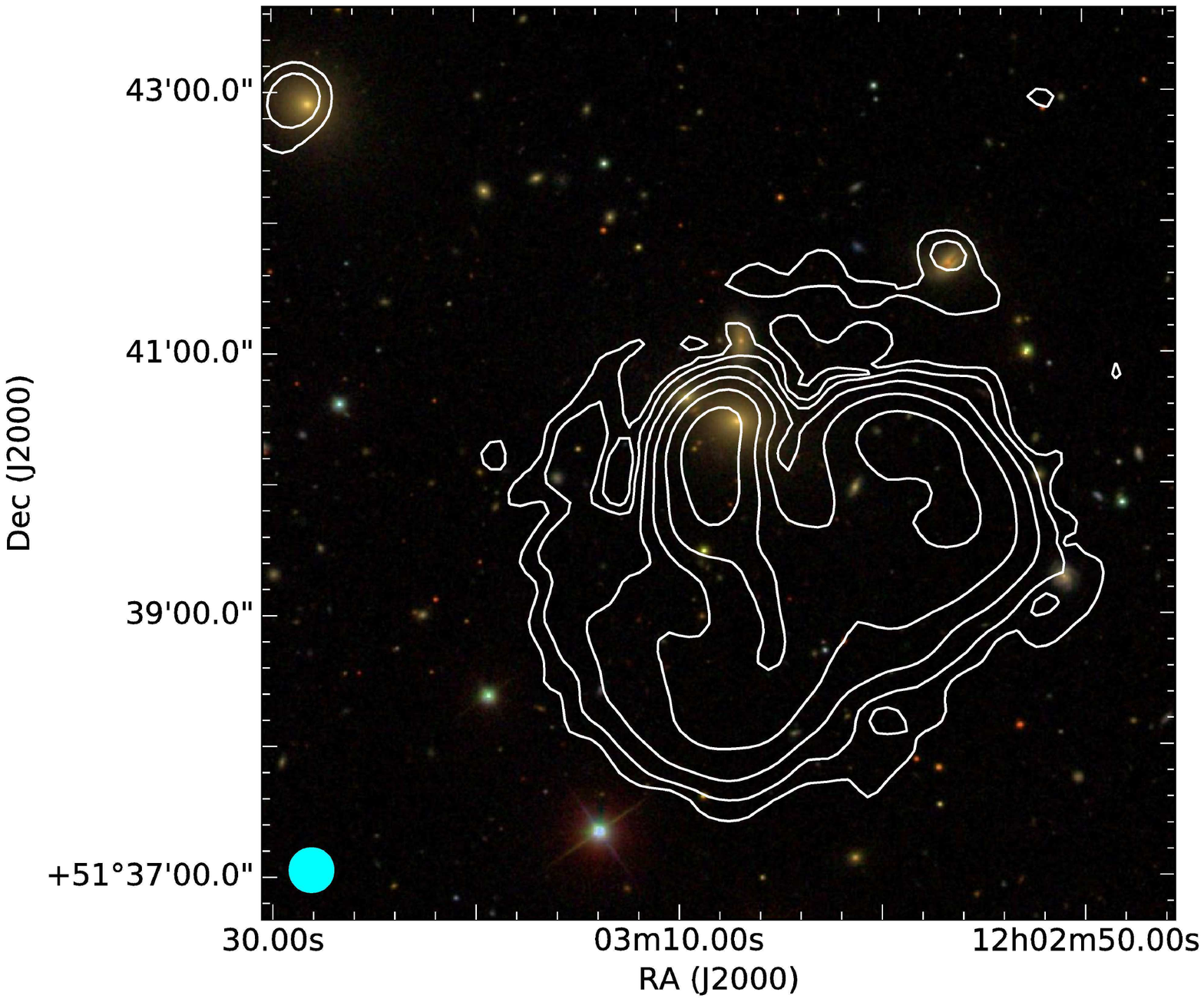}
    \includegraphics[width=0.49\textwidth]{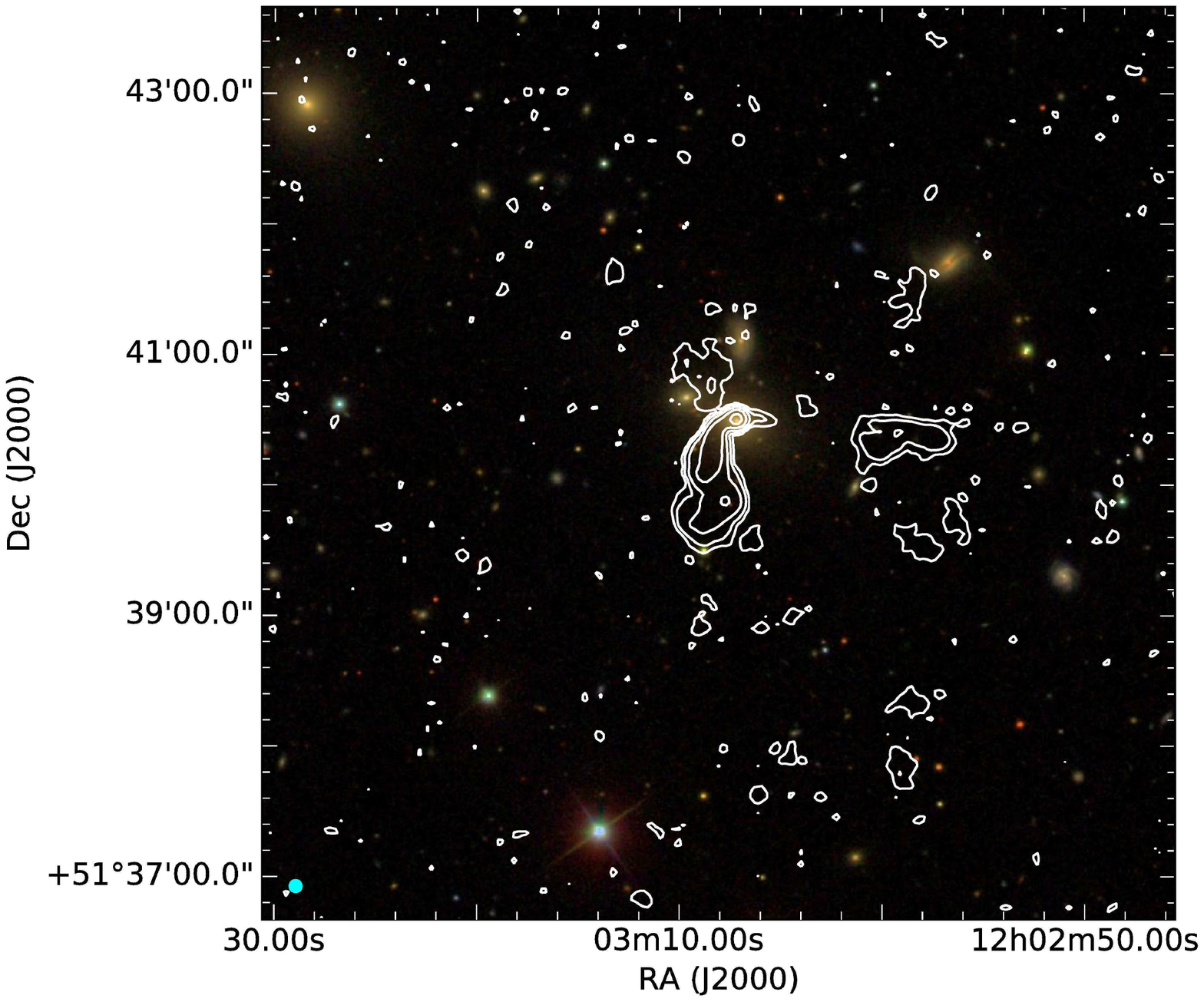}
    \includegraphics[width=0.49\textwidth]{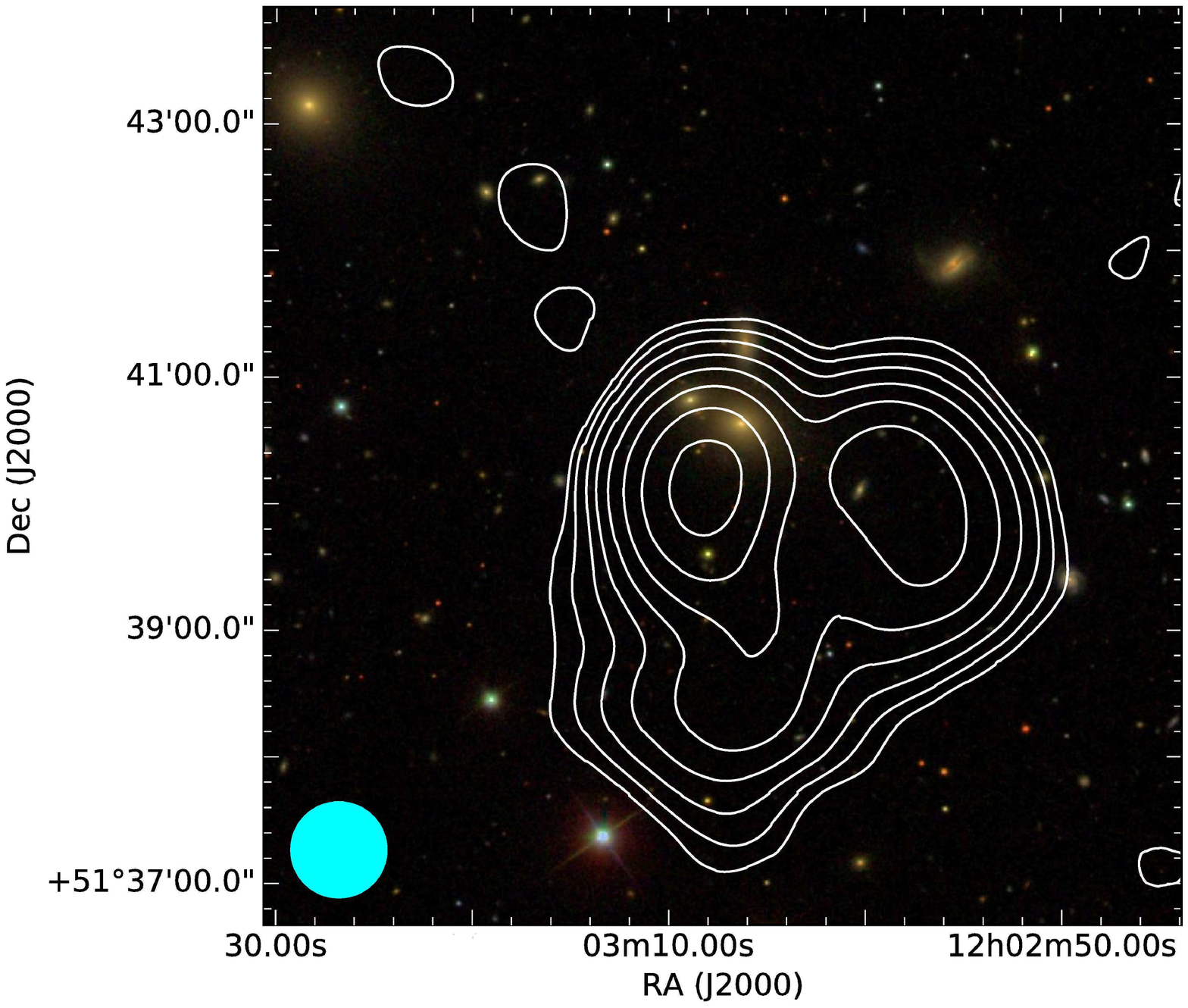}
\caption{\label{fig:hcg60}Contour maps of the radio emission of HCG\,60 from different surveys overlaid on an optical \textit{gri} RGB-composite image from the SDSS. The contour levels start at $3\sigma$ noise level and increase by a factor of $2\!\sqrt{2}$. The beam area is represented by a cyan circle in the lower left corner of the image. The size of the plotted cutouts is 7$\times$7\,arcmin$^{2}$.
\textbf{Upper left:} LoTSS, 6 arcsecs resolution, $\sigma$ noise of 0.4 mJy/beam; 
\textbf{upper right:} LoTSS, 20 arcsecs resolution, $\sigma$ noise of 0.4 mJy/beam; 
\textbf{lower left:} FIRST, 5 arcsecs resolution, $\sigma$ noise of 0.15 mJy/beam; 
\textbf{and lower right:} NVSS, 45 arcsecs resolution, $\sigma$ noise of 0.45 mJy/beam.}     
\end{figure*}

\begin{figure*} 
\centering
        \includegraphics[width=0.49\textwidth]{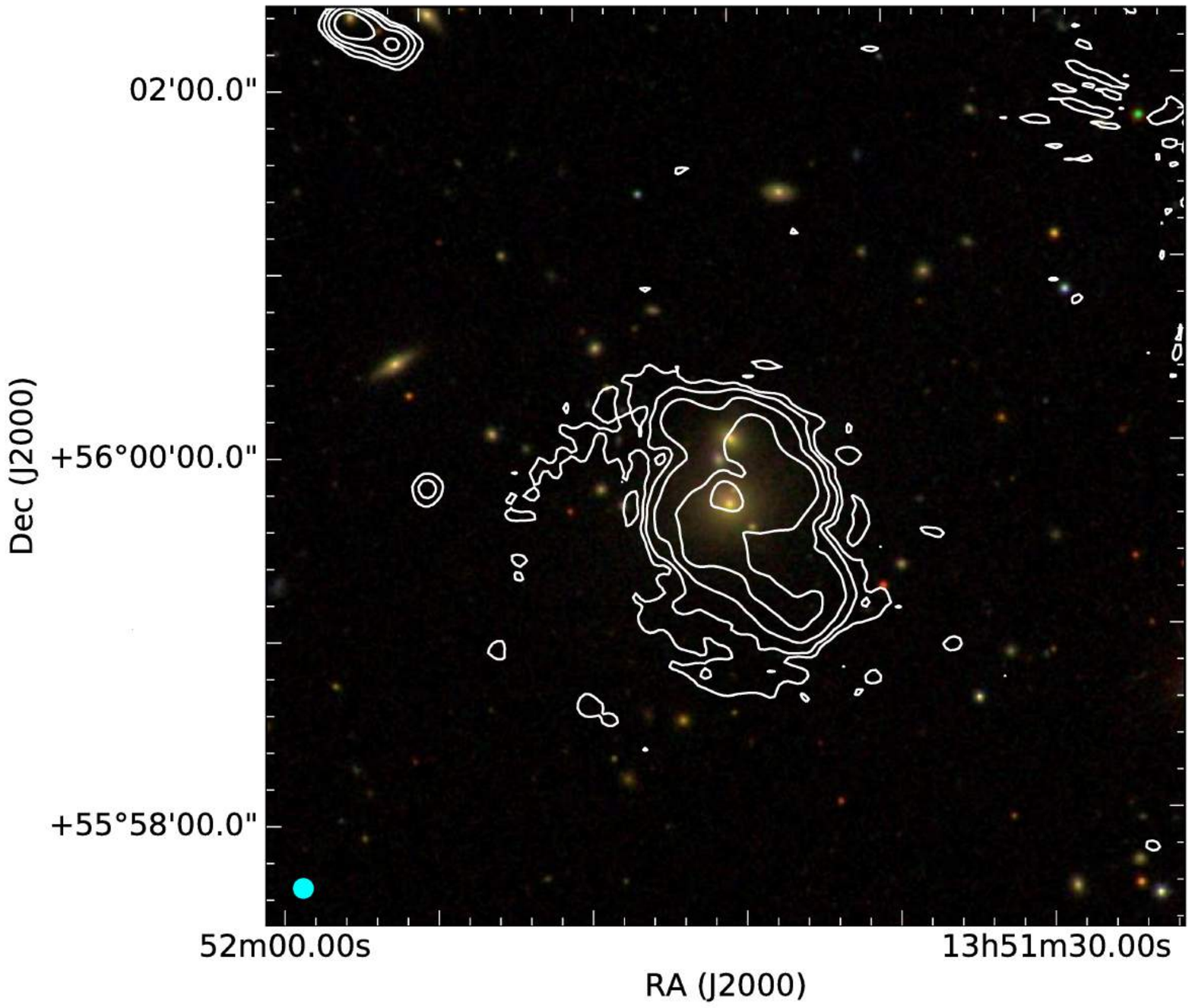}
    \includegraphics[width=0.49\textwidth]{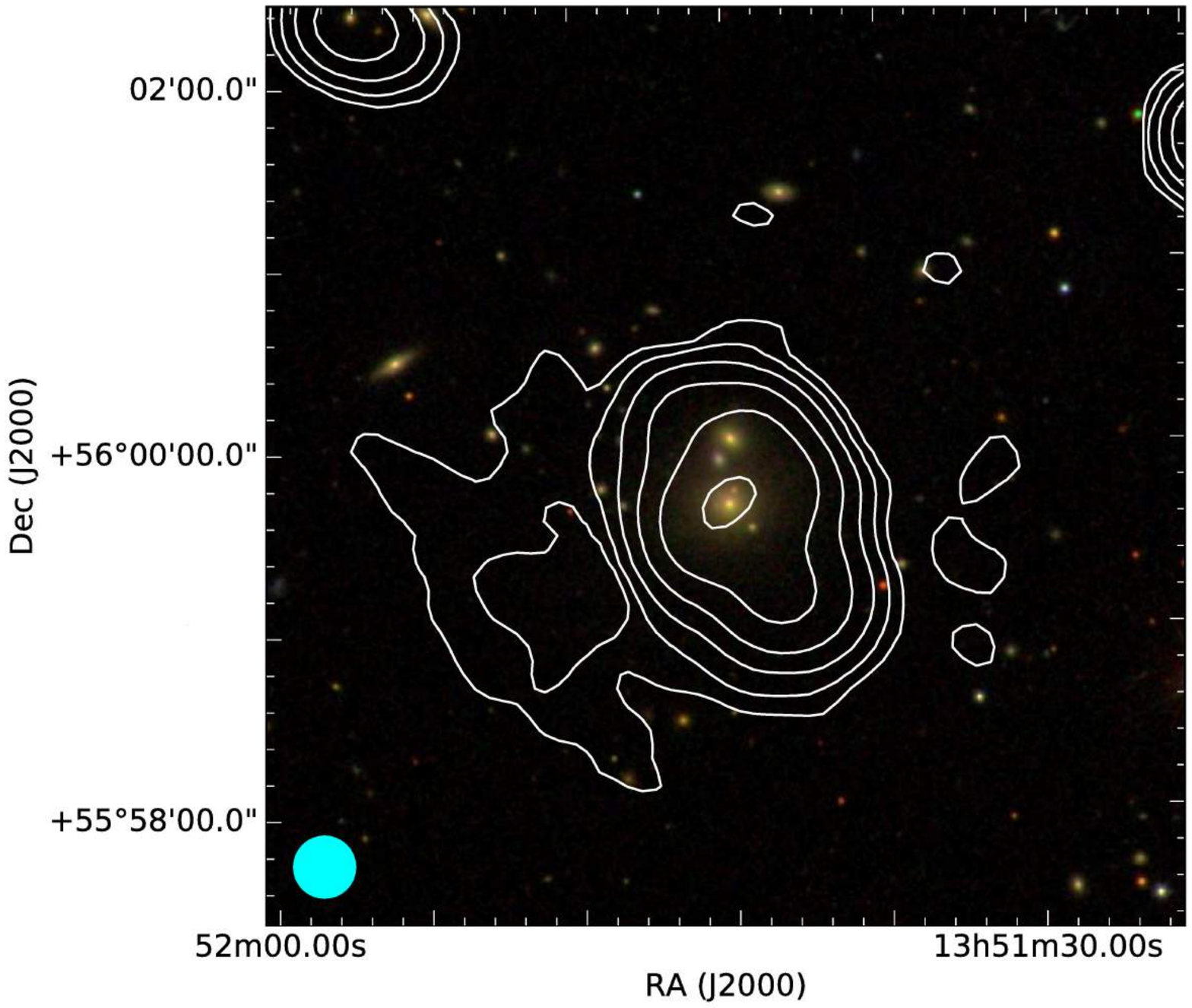}
    \includegraphics[width=0.49\textwidth]{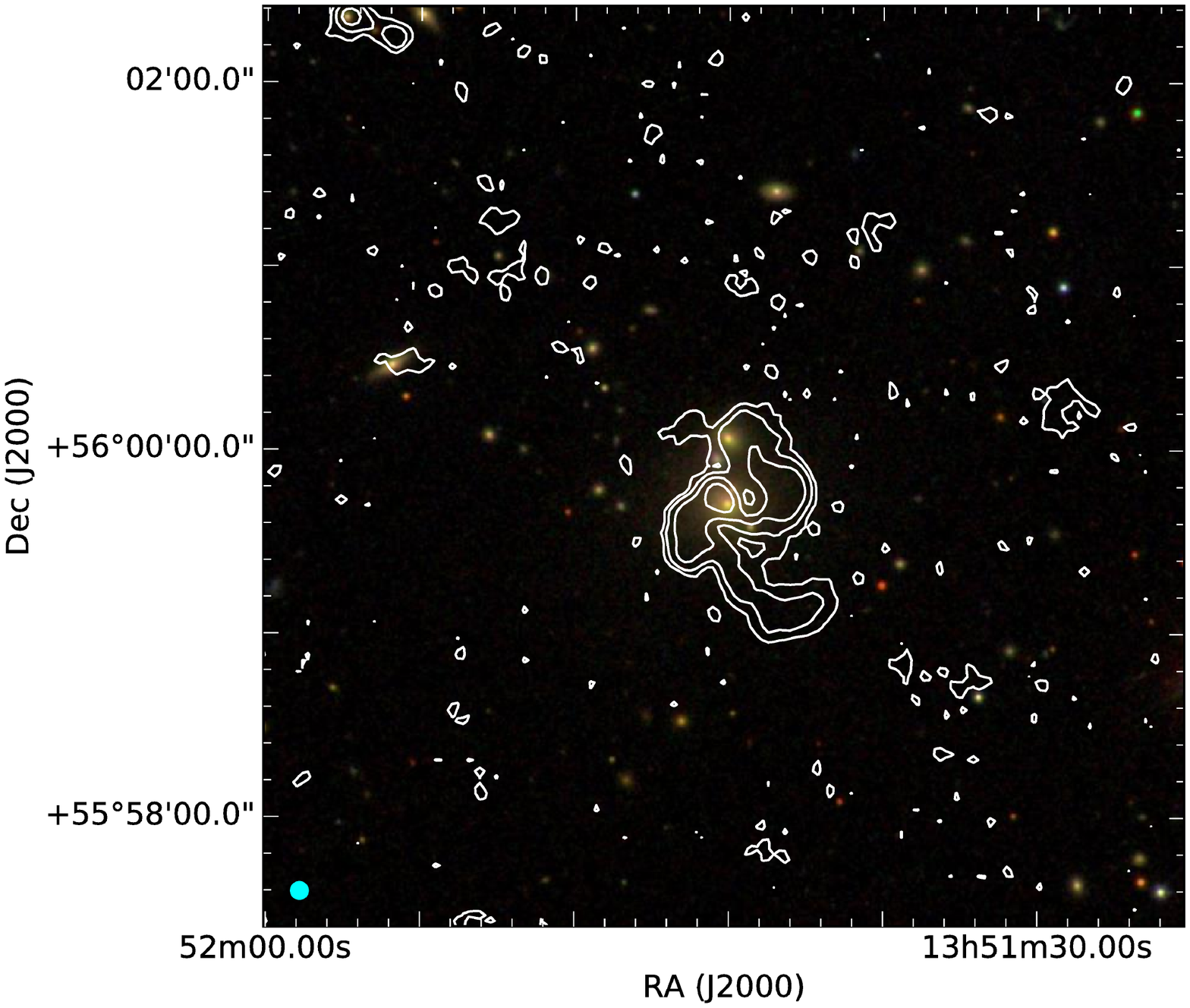}
    \includegraphics[width=0.49\textwidth]{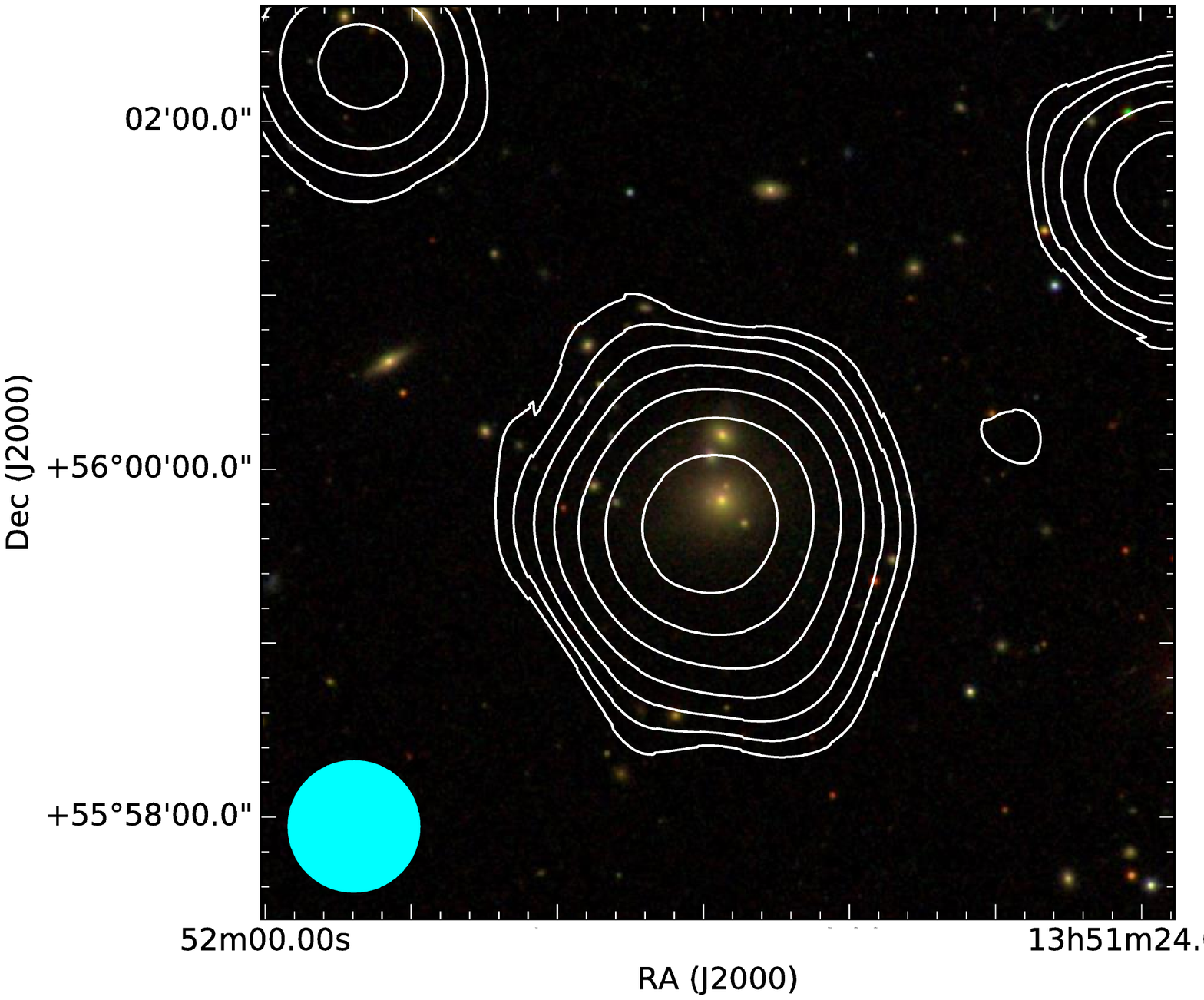}
\caption{\label{fig:mlcg24} As in Fig.~\ref{fig:hcg60}, for MLCG\,24. The size of the plotted cutouts is 5$\times$5\,arcmin$^{2}$.}    
\end{figure*}

HCG\,60 and MLCG\,24 are systems that are both suspected to be parts of larger structures. The former  (presented in Fig.~\ref{fig:hcg60}) was already identified as a central region of the Abell\,1452 cluster \citep{abell58}; in the case of the latter, no parent structure has been identified so far, but there are hints that it could form a loose group with some of the nearby field objects \citep{mlcg}. The other similarity is that these systems seem to be wrapped around powerful radio galaxies. The well-known HCG\,60 hosts a wide angle tail galaxy (WAT), studied in detail by \citet{miley77}, \citet{rudnick76}, and \citet{jagers87}: its unusual shape results from its movement through the intergalactic medium (IGM). MLCG\,24 was studied more scarcely: analysis of its radio emission (at 1400\,MHz) was provided by \citet{berger16}, who described the unusual radio structure found inside. The high-resolution LoTSS image of this group (Fig.~\ref{fig:mlcg24}, upper left panel) reveals a well-defined radio structure, similar in its shape to an inverted letter "S"; it is partially visible in the FIRST data (lower left panel), but much less pronounced and without the eastern "tail" because of the sparse $(u,v)$ coverage of this survey. Both in the NVSS (lower right panel) and in lower resolution LoTSS data (upper right panel), much of the detail is lost, albeit the latter data reveal  hints of the S-shaped structure. The total flux density of $926 \pm 46 $\,mJy is somewhat lower, but still consistent with the measurement from the Tata Institute for Fundamental Research Giant Metrewave Telecope Sky Survey Alternative Data Release (TGSS ADR) \citep{tgssadr}, where it was equal to $1085 \pm 109$\,mJy.

In case of HCG\,60, the low-frequency structure is similar to what has been found at 1400\,MHz;  the important difference is that much more, and more detailed, emission can be seen. In particular, LoTSS high-resolution data (Fig.~\ref{fig:hcg60}, upper left panel) show almost the same extent as the NVSS map (lower right panel of the same figure), but the structure of the radio galaxy can be easily disentangled from its envelope. The shape of the radio structure suggests that most of the observed emission comes from two swept-up lobes that are connected to the central engine by symmetrical jets. The eastern lobe bends closer to the AGN and is more luminous, signifying that this is the colliding side. It is worth noticing that the radio-loud part of this system is accompanied by much weaker, extended structures, which are presumably lobes and radio-emitting IGM. Data from LoTSS deal well with revealing these areas; the negligible difference between the flux density of the smoothed and non-smoothed maps (see Table~\ref{tab:fluxes}) means that the 6 arcsec resolution does not lead to over-resolving the structure of the group, while simultaneously allowing us to describe its details.These weaker entities are not represented in the TGSS ADR, where the total flux density of HCG\,60 is $1810 \pm 180$\,mJy; that is, around two-thirds of what was detected in the LoTSS.

\subsubsection{Typical case of a non-AGN-dominated, isolated group: MLCG\,41}
\label{results_mlcg41}

\begin{figure*} 
\centering
        \includegraphics[width=0.49\textwidth]{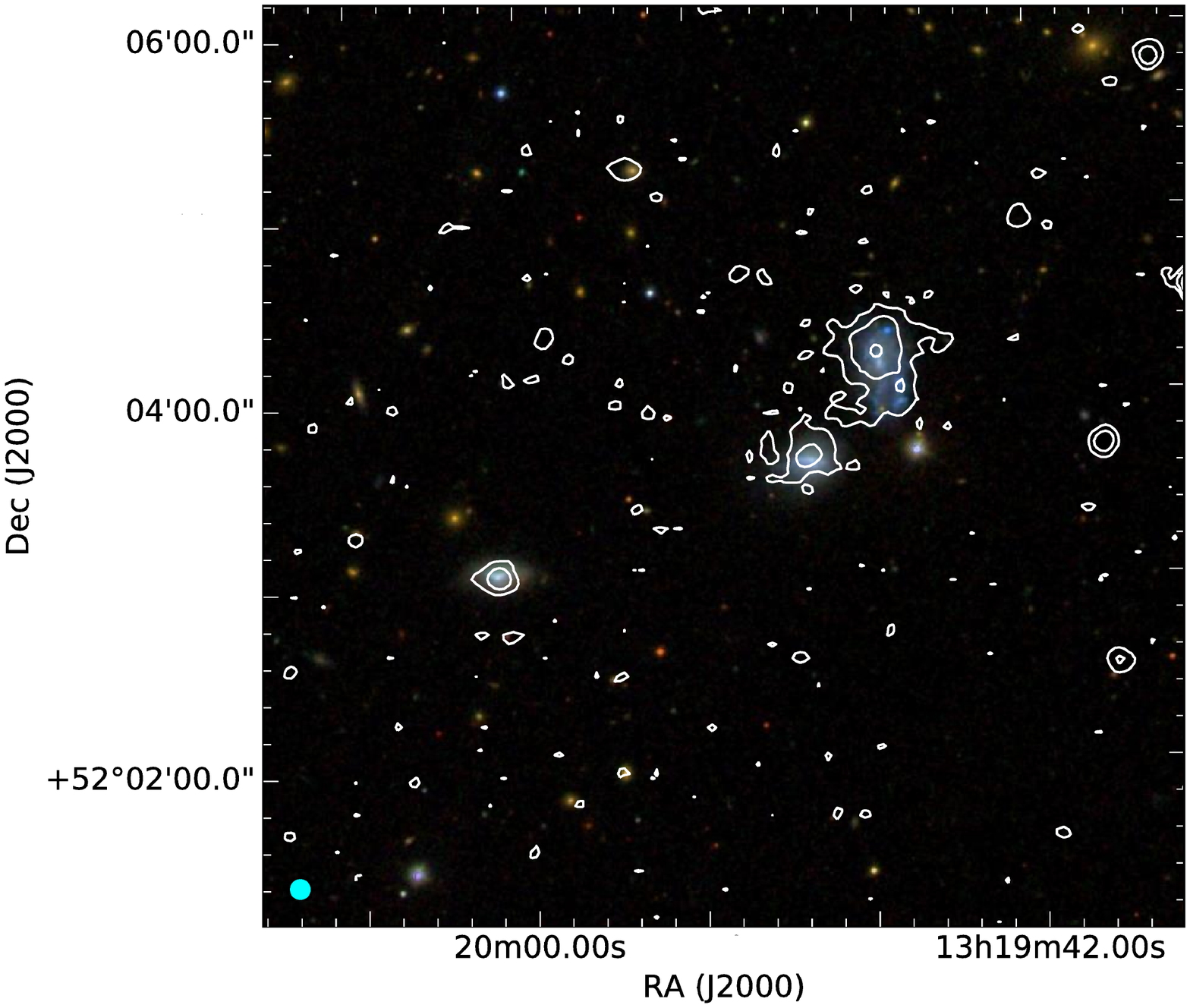}
    \includegraphics[width=0.49\textwidth]{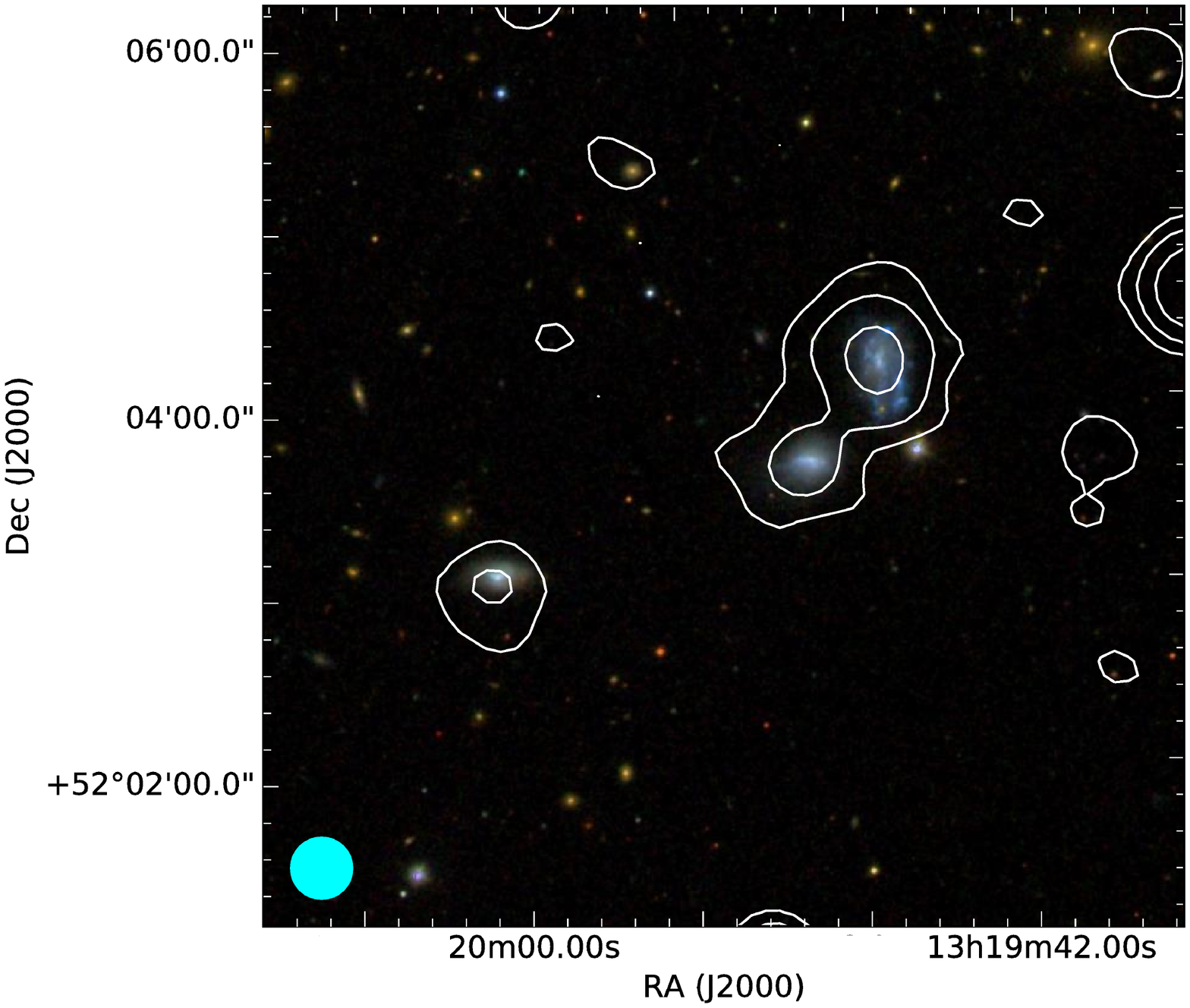}
    \includegraphics[width=0.49\textwidth]{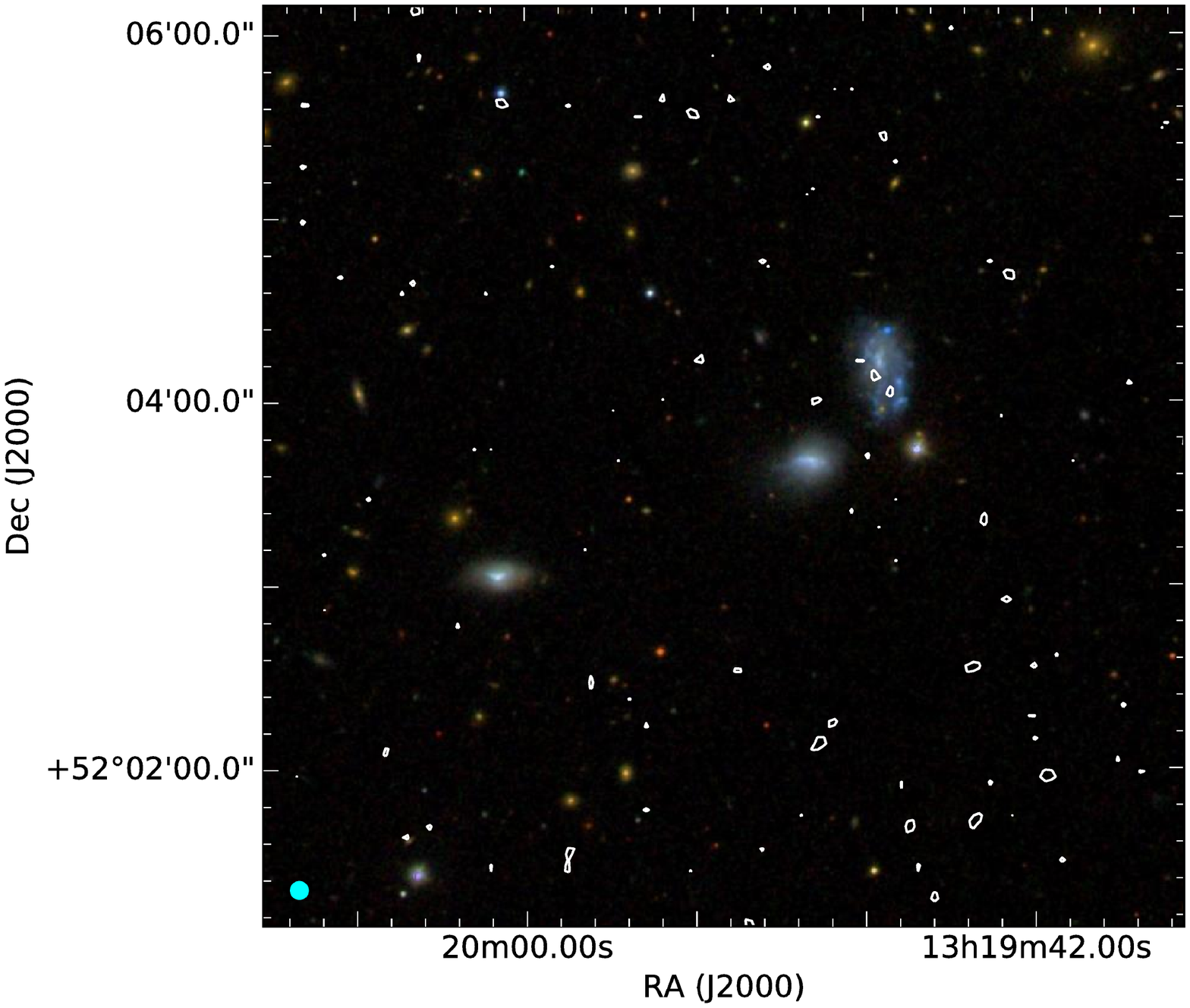}
    \includegraphics[width=0.49\textwidth]{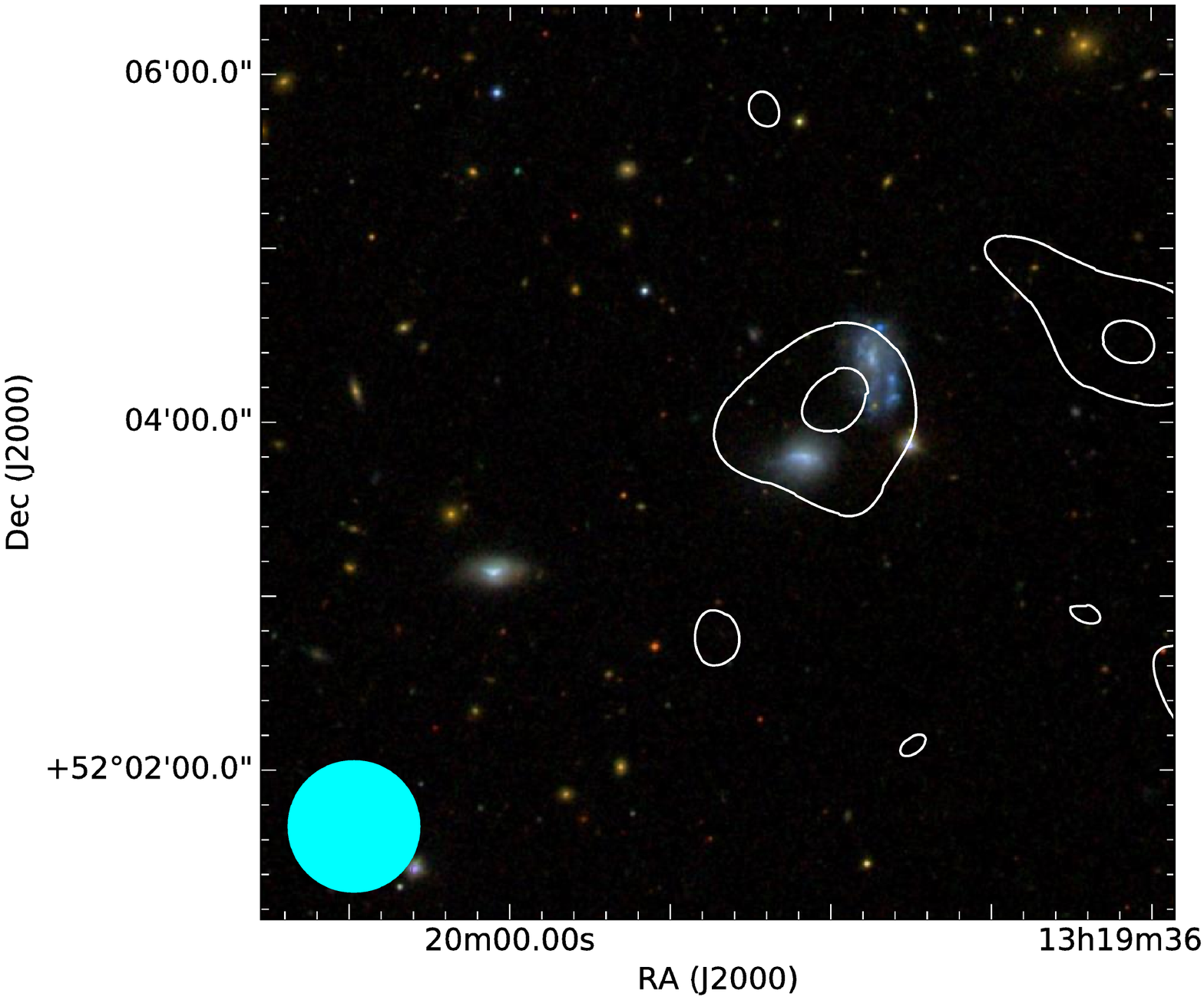}
\caption{\label{fig:mlcg41} As in Fig.~\ref{fig:hcg60}, for MLCG\,41. The size of the plotted cutouts is 5$\times$5\,arcmin$^{2}$.
}    
\end{figure*}

MLCG\,41 can be regarded as a model galaxy group, at least for the sake of discussing its radio picture. It is a moderately dense galaxy system and has a relatively small angular size of 2 arcmins. It features only three galaxies, none of which show hints of strong AGN activity \citep{mlcg}. The central object of this system is a blue, irregular galaxy that is in the process of colliding with the southern edge-on spiral member. Signs of interaction can be readily spotted from the optical images (Fig.~\ref{fig:mlcg41}): strains of loose stars protruding from the irregular galaxy are elongated towards the spiral galaxy. At 1400\,MHz, the NVSS image (Fig.~\ref{fig:mlcg41}, lower right panel) seems to show a connection between the colliding objects, with the maximum of the radio emission in intergalactic space; however, the signal-to-noise ratio of this structure is low, as it barely exceeds the declared $5\sigma$ level of this map. Also, presence of extended radio structures with no optical counterparts puts the reality of the supposed bridge in doubt. The FIRST data (Fig.~\ref{fig:mlcg41}, lower left panel) are not helpful in this particular case: no certain detection of radio emission was made. Whereas the data suggest that most of the emission that NVSS recovered is due to weak, extended structures, it is impossible to discuss neither their origin nor morphology. At 150\,MHz, LoTSS data reveal a connection between the two galaxies; moreover, the third member of the group, albeit not connected to the rest, is also a radio-emitting galaxy, whereas it was too weak to be detected in the higher frequency data.
The flux density measured for the intergalactic filaments is equal to 3.8\,mJy, which accounts for approximately 25\% of the total flux density (15.4\,mJy). This system was not detected in the TGSS ADR (there is only noise at the position of this group), so no comparison is possible.

\subsubsection{Distant and angularly small object: MLCG\,1374, the 'Wristwatch'}
\label{results_mlcg1374}

\begin{figure*}
\centering
        \includegraphics[width=0.49\textwidth]{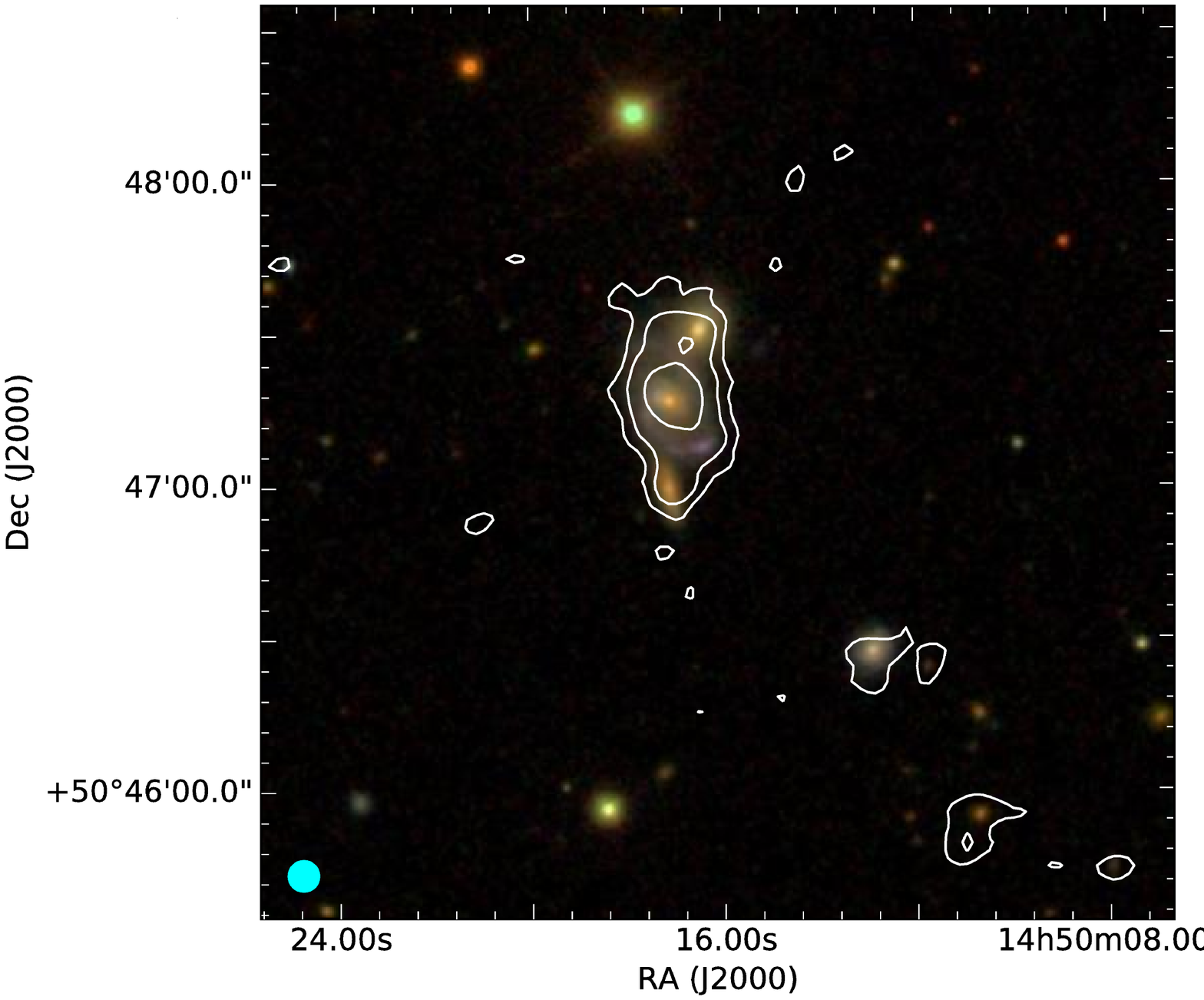}
    \includegraphics[width=0.49\textwidth]{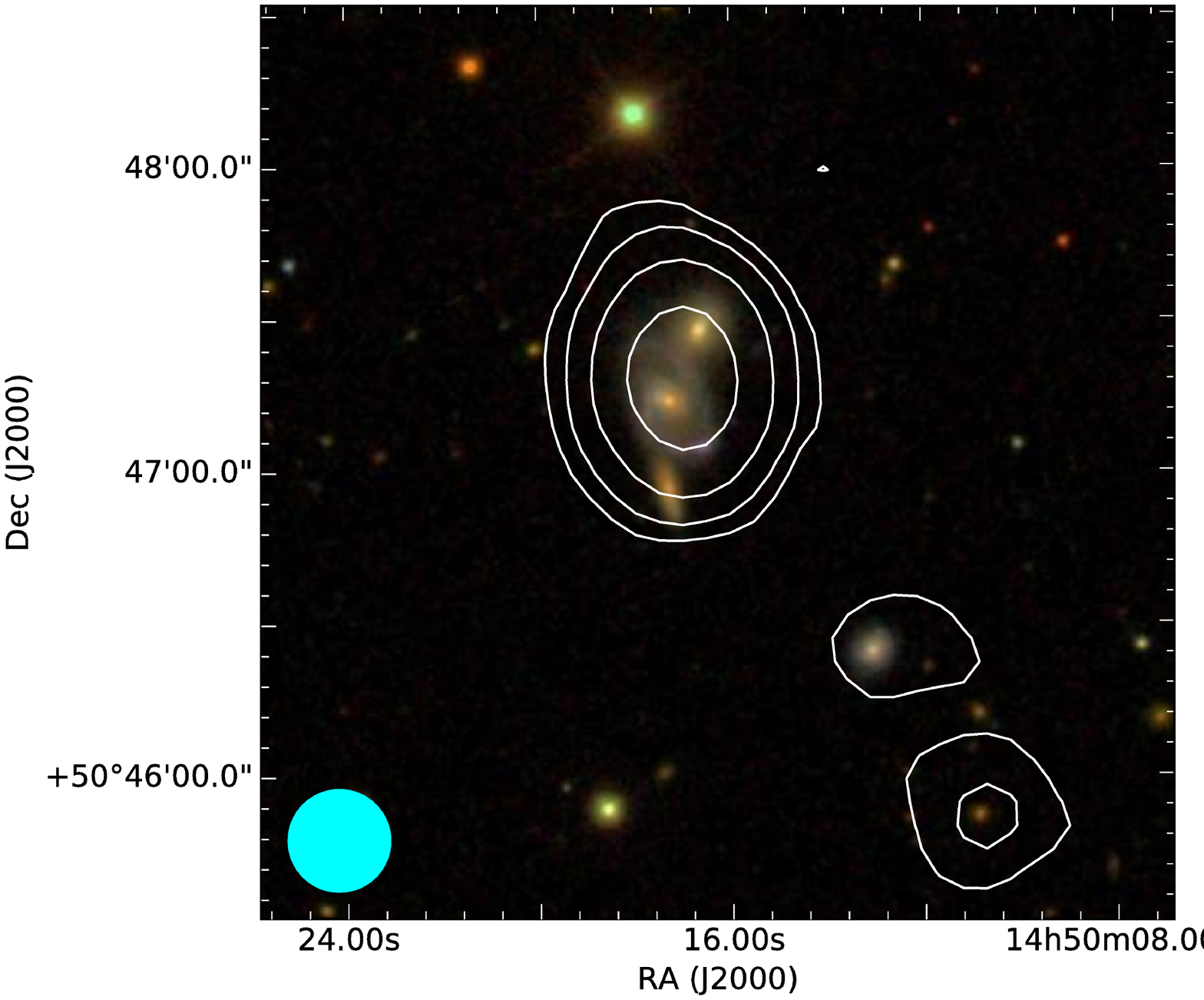}
    \includegraphics[width=0.49\textwidth]{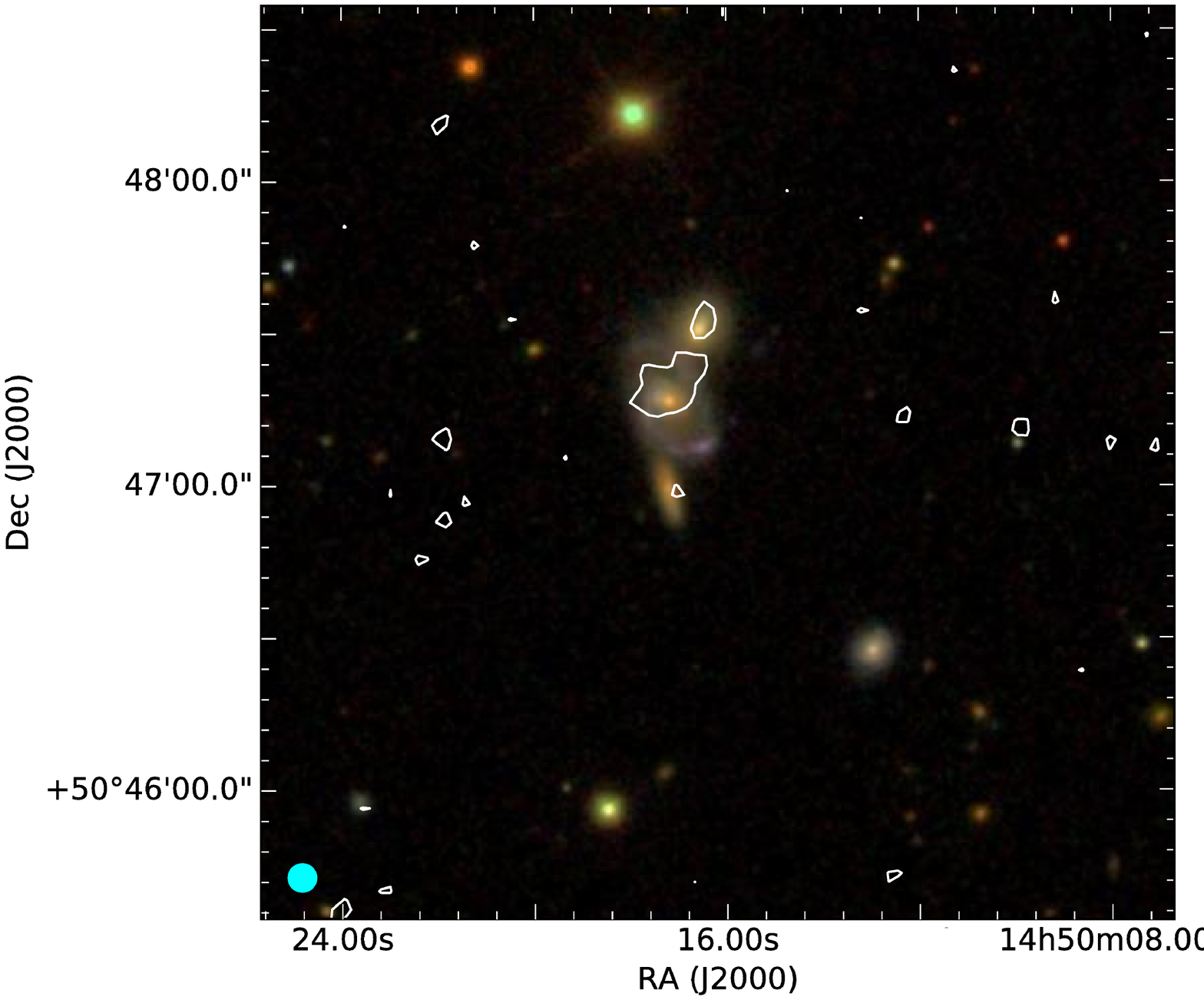}
    \includegraphics[width=0.49\textwidth]{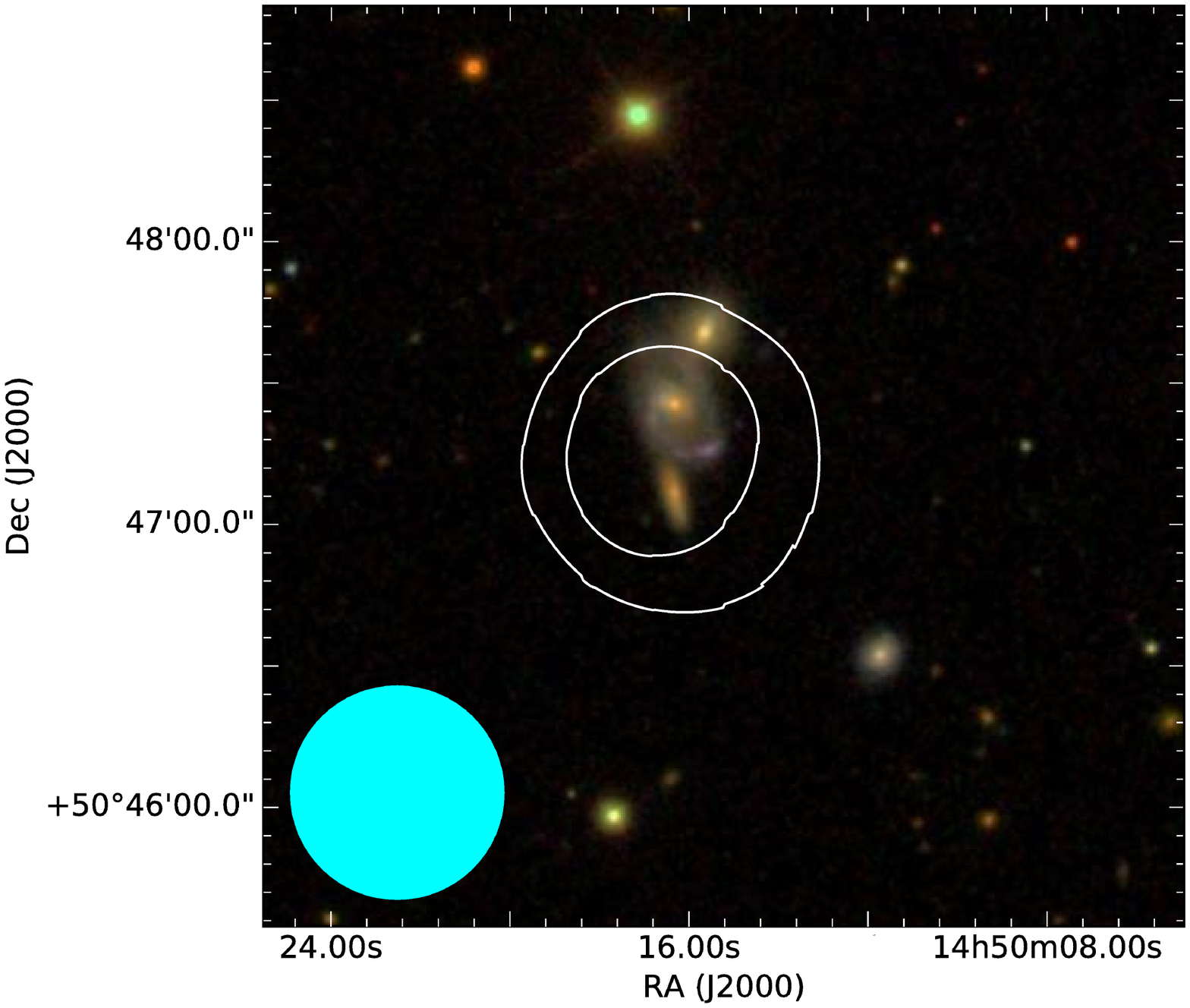}
\caption{\label{fig:mlcg1374} As in Fig.~\ref{fig:hcg60}, for MLCG\,1374.  The size of the plotted cutouts is 3$\times$3\,arcmin$^{2}$.
}      
\end{figure*}

MLCG\,1374 is a system that seems to be, similar to the famous "Taffy" pairs of galaxies \citep{condon93,condon02}, not a very interesting target while viewing its optical image only. This angularly small quartet is located nearly 300\,Mpc from the Milky Way (according to \citealt{RC3}), and three of its galaxies form a nearly-straight line that has a total extent of less than a single arcmin. These star-forming, spiral galaxies are visibly colliding with each other.
The 1400\,MHz-radio data do not reveal interesting features. Whereas only the central object is visible in the FIRST data (Fig.~\ref{fig:mlcg1374}, lower left panel) with hints of some extensions (although owing to the presence of artefacts they are ambiguous and non-convincing),  the NVSS (lower right) shows just a single source of radio emission. The situation gets strikingly different while viewing this system in the LoTSS. The 20 arcsec map (upper right panel) does, similar to the NVSS, show just a single blob of emission (albeit weak extensions could be seen), but the 6 arcsec map (upper left panel) reveals the reason for its namesake. The central galaxy, supposedly an AGN (or a star-forming nuclei), acts like a dial of a wristwatch, wherein the two companions, each clearly showing extended, disc emission, form "belts"; as all three galaxies have consistent redshifts \citep{mlcg}, this means that the radio-emitting structure is a genuine radio envelope of this system. Subtraction of the point sources suggests that the flux of the extended structure is as high as 18\,mJy, which is more than 65\% of the total flux density. Exact decomposition into galactic discs
 and intergalactic filaments is, however, impossible at the moment. Similar to MLCG\,41, there is no detection in TGSS ADR.

\onecolumn
\begin{longtable}[l]{lccccccccc}

\caption{\label{tab:cat_groups} Sky position, redshift, flux densities, type of the emission, number of members, and their overall abundance for HCG and MLCG systems detected in the LoTSS.}\\
\hline
\hline
Name    &       Subset  &       RA(deg) &       Dec(deg)        &       z       &       Si6(mJy)        &       Si6err(mJy)     &       Type    &       Nrcomp  &       Nremitt \\
\hline
\endfirsthead
\caption{continued.}\\
\hline\hline
Name    &       Subset  &       RA(deg) &       Dec(deg)        &       z       &       Si6(mJy)        &       Si6err(mJy)     &       Type    &       Nrcomp  &       Nremitt \\
\hline
\endhead
\hline
\endfoot
HCG56           &       C       &       173.14723       &       52.94724        &       0.0268  &       ?       &       ?       &       E*      &       5       &       5       \\
HCG60           &       E       &       180.79004       &       51.67817        &       0.0625  & 2408.7& 90.5    &       E       &       4       &       3       \\
MLCG8           &       E       &       223.76506       &       54.32933        &       0.0996  &       0.9     &       0.1     &       G       &       3       &       1       \\
MLCG10          &       C       &       188.8597        &       47.73041        &       0.0309  &       5.9     &       0.3     &       G       &       4       &       1       \\
MLCG23          &       C       &       207.7261        &       56.03778        &       0.0687  &       2.5     &       0.2     &       G       &       3       &       1       \\
MLCG24          &       E       &       207.92557       &       55.99534        &       0.0691  & 925.7   & 46.3  &       E       &       3       &       3       \\
MLCG28          &       E       &       211.8439        &       54.00246        &       0.1162  &       ?       &       ?       &       G**     &       3       &       1       \\
MLCG39          &       C       &       188.82079       &       47.29997        &       0.031   &       1.1     &       0.1     &       G       &       3       &       2       \\
MLCG41          &       C       &       199.9581        &       52.06146        &       0.0154  &       15.4&   0.6     &       E       &       3       &       3       \\
MLCG44          &       C       &       229.85564       &       49.50135        &       0.0373  &       0.7     &       0.1     &       G       &       3       &       1       \\
MLCG45          &       C       &       198.9381        &       55.7888         &       0.0811  &       6.6     &       0.2     &       E**     &       3       &       3       \\
MLCG51          &       C       &       197.83965       &       47.52277        &       0.0291  &       4.9     &       0.3     &       G       &       3       &       1       \\
MLCG118         &       C       &       223.53822       &       46.99274        &       0.0873  &       ?       &       ?       &       G       &       3       &       1       \\
MLCG176         &       C       &       216.32787       &       47.25048        &       0.0714  &       8.1     &       0.4     &       E       &       4       &       2       \\
MLCG177         &       E       &       209.63638       &       49.53931        &       0.1043  &       0.4     &       0.1     &       G       &       3       &       1       \\
MLCG1108        &       C       &       171.63628       &       55.57245        &       0.0471  &       1.5     &       0.1     &       G       &       3       &       1       \\
MLCG1116        &       C       &       163.54324       &       54.82738        &       0.0728  &       ?       &       ?       &       G**     &       3       &       1       \\
MLCG1117        &       C       &       163.40237       &       54.86793        &       0.0716  &       15.7&   0.8     &       G       &       3       &       1       \\
MLCG1121        &       C       &       170.7665        &       56.0237         &       0.0592  &       ?       &       ?       &       G*      &       4       &       2       \\
MLCG1130        &       C       &       178.69844       &       52.82193        &       0.0677  &       0.6     &       0.1     &       G       &       3       &       1       \\
MLCG1131        &       C       &       162.4909        &       51.84148        &       0.0241  &       ?       &       ?       &       N       &       3       &       ?       \\
MLCG1138        &       C       &       166.99965       &       55.88076        &       0.0505  &       ?       &       ?       &       G       &       3       &       1       \\
MLCG1149        &       C       &       164.99811       &       50.05664        &       0.0236  &       ?       &       ?       &       G       &       3       &       2       \\
MLCG1167        &       C       &       177.87375       &       54.22488        &       0.0662  &       5.1     &       0.3     &       G       &       3       &       1       \\
MLCG1186        &       C       &       172.4721        &       50.13337        &       0.0471  &       2.4     &       0.1     &       G       &       3       &       2       \\
MLCG1191        &       C       &       177.64488       &       50.52897        &       0.023   &       2.6     &       0.1     &       G       &       3       &       2       \\
MLCG1192        &       E       &       197.03194       &       50.07972        &       0.055   &       4.1     &       0.2     &       G       &       4       &       2       \\
MLCG1193        &       C       &       199.12987       &       49.91392        &       0.0487  &       ?       &       ?       &       G       &       3       &       3       \\
MLCG1195        &       E       &       217.10788       &       45.96946        &       0.0738  &       ?       &       ?       &       E       &       3       &       2       \\
MLCG1202        &       E       &       173.06662       &       55.96308        &       0.0515  & 538.4   & 25.6  &       E       &       5       &       4       \\
MLCG1204        &       C       &       178.0392        &       53.3167         &       0.0558  &       14.0&   0.5     &       E       &       3       &       3       \\
MLCG1208        &       C       &       178.55063       &       52.69115        &       0.0646  &       ?       &       ?       &       N       &       3       &       ?       \\
MLCG1216        &       C       &       208.16902       &       48.7993         &       0.0683  &       0.9     &       0.1     &       G       &       3       &       2       \\
MLCG1218        &       C       &       187.72652       &       51.34129        &       0.0441  &       4.2     &       0.2     &       E**     &       4       &       4       \\
MLCG1225        &       C       &       177.81476       &       56.39917        &       0.0182  &       3.7     &       0.2     &       G       &       3       &       1       \\
MLCG1227        &       E       &       177.00093       &       54.6387         &       0.0591  &       8.4     &       0.4     &       G       &       4       &       1       \\
MLCG1246        &       C       &       193.4826        &       48.78237        &       0.0378  &       9.2     &       0.5     &       G       &       4       &       1       \\
MLCG1258        &       C       &       214.85023       &       45.90078        &       0.0265  &       3.4     &       0.2     &       G       &       3       &       1       \\
MLCG1259        &       E       &       172.29884       &       54.11736        &       0.0684  &       0.6     &       0.1     &       G       &       3       &       1       \\
MLCG1262        &       C       &       169.28462       &       54.91743        &       0.1369  &       3.0     &       0.2     &       G       &       5       &       2       \\
MLCG1264        &       C       &       164.04887       &       46.88068        &       0.0286  &       7.5     &       0.3     &       G       &       3       &       3       \\
MLCG1266        &       C       &       169.86694       &       54.36284        &       0.07    &       0.3     &       0.1     &       G       &       3       &       1       \\
MLCG1267        &       C       &       224.93533       &       47.94507        &       0.0312  &       ?       &       ?       &       N       &       3       &       1       \\
MLCG1270        &       C       &       181.95894       &       51.60939        &       0.0878  &       1.9     &       0.1     &       G       &       3       &       1       \\
MLCG1276        &       C       &       184.66737       &       48.87935        &       0.0452  &       0.5     &       0.1     &       G       &       3       &       1       \\
MLCG1278        &       C       &       221.20508       &       48.64876        &       0.0496  &       24.8&   0.2     &       G       &       3       &       2       \\
MLCG1283        &       C       &       186.95001       &       50.38782        &       0.04    &       14.5&   0.6     &       G       &       3       &       2       \\
MLCG1284        &       C       &       185.42297       &       55.78717        &       0.0343  &       1.2     &       0.1     &       G       &       3       &       1       \\
MLCG1296        &       E       &       173.93216       &       49.16663        &       0.0353  &       1.1     &       0.1     &       G       &       3       &       1       \\
MLCG1299        &       C       &       177.92174       &       54.97033        &       0.0601  &       ?       &       ?       &       G       &       3       &       ?       \\
MLCG1309        &       E       &       179.18962       &       55.0303         &       0.0638  &       1.7     &       0.1     &       G       &       3       &       2       \\
MLCG1310        &       C       &       166.61151       &       48.63585        &       0.0246  &       24.3&   1.2     &       G       &       3       &       2       \\
MLCG1322        &       C       &       226.91379       &       52.83156        &       0.1055  &       1.1     &       0.1     &       G       &       3       &       1       \\
MLCG1323        &       C       &       219.6744        &       56.0012         &       0.0428  &       1.2     &       0.1     &       G       &       3       &       2       \\
MLCG1324        &       C       &       220.31392       &       55.84566        &       0.0505  &       5.0     &       0.3     &       G       &       3       &       1       \\
MLCG1326        &       E       &       170.00888       &       47.2397         &       0.1097  &       ?       &       ?       &       E**     &       3       &       ?       \\
MLCG1332        &       E       &       168.72821       &       54.4709         &       0.0675  &       8.5     &       0.4     &       G       &       4       &       1       \\
MLCG1334        &       C       &       171.6179        &       54.78723        &       0.0471  &       69.5&   3.4     &       E**     &       5       &       3       \\
MLCG1335        &       C       &       173.90157       &       54.94856        &       0.0193  &       0.3     &       0.1     &       G       &       3       &       1       \\
MLCG1349        &       C       &       185.15677       &       47.64976        &       0.0588  &       8.6     &       0.4     &       G       &       3       &       2       \\
MLCG1350        &       E       &       225.32376       &       47.27852        &       0.0867  &       8.0     &       0.5     &       E       &       3       &       1       \\
MLCG1361        &       C       &       174.50853       &       47.4835         &       0.0337  &       1.5     &       0.1     &       G       &       3       &       1       \\
MLCG1362        &       C       &       170.96927       &       47.19486        &       0.0529  &       4.5     &       0.2     &       G       &       5       &       1       \\
MLCG1364        &       C       &       205.93755       &       55.63378        &       0.0682  &       4.0     &       0.2     &       G       &       5       &       3       \\
MLCG1365        &       E       &       205.80719       &       55.62277        &       0.0663  &       19.1&   0.9     &       E       &       4       &       2       \\
MLCG1370        &       C       &       205.91328       &       55.37853        &       0.0676  &       1.0     &       0.1     &       G       &       3       &       1       \\
MLCG1371        &       C       &       181.22063       &       54.23176        &       0.0503  &       5.3     &       0.3     &       G       &       3       &       1       \\
MLCG1372        &       C       &       213.35889       &       54.73987        &       0.0834  &       5.8     &       0.3     &       G       &       3       &       3       \\
MLCG1373        &       C       &       221.21976       &       51.34578        &       0.0897  &       1.1     &       0.1     &       G       &       4       &       2       \\
MLCG1374        &       C       &       222.5684        &       50.79213        &       0.0697  &       26.9&   1.3     &       E       &       4       &       4       \\
MLCG1375        &       C       &       204.5809        &       56.40196        &       0.0523  &       2.4     &       0.1     &       G       &       3       &       1       \\
MLCG1379        &       C       &       218.30403       &       52.96314        &       0.0473  &       10.9&   0.5     &       G       &       3       &       2       \\
MLCG1380        &       C       &       178.97305       &       48.28494        &       0.0318  &       2.8     &       0.1     &       G       &       3       &       2       \\
MLCG1495        &       E       &       219.74133       &       46.68961        &       0.0364  &       3.0     &       0.2     &       E**     &       3       &       1       \\
MLCG1562        &       C       &       174.78572       &       55.66449        &       0.0612  & 1358.2& 67.9    &       E       &       3       &       2       \\
MLCG1563        &       E       &       203.94296       &       47.43361        &       0.0645  &       14.2&   0.7     &       G       &       3       &       1       \\
MLCG1564        &       C       &       176.8391        &       55.73018        &       0.0515  &       5.3     &       0.2     &       G       &       6       &       2       \\
MLCG1565        &       C       &       176.90718       &       55.71792        &       0.0491  &       2.3     &       0.2     &       G       &       3       &       1       \\
MLCG1567        &       E       &       173.7054        &       49.07764        &       0.0333  &       12.9&   6.4     &       G       &       4       &       1       \\
MLCG1568        &       C       &       162.9318        &       51.02216        &       0.025   &       20.0&   1.0     &       G       &       3       &       2       \\
\hline
\end{longtable}
\tablefoot{
Column descriptions:\\
Name: name of the system;\\
Subset: either \textbf{I}solated, or \textbf{C}ontained;\\
RA, Dec: sky position (in degrees) of the system;\\
z: spectroscopic redshift from \citet{mlcg};\\
Si6: integrated (measured by \textsc{blobcat}) flux density of the group;\\
Si6err: uncertainty of the latter;\\
Type: emission type; \textbf{G}alactic, or \textbf{E}nvelope;\\
        *: visible contamination from background sources;\\
    ** ambiguous detection (e.g. low signal-to-noise ratio);\\
Nrcomp: number of members from \citet{mlcg};\\
Nremitt: number of emitting components.\\
}

\begin{longtable}[l]{lcccccc}

\caption{\label{tab:cat_gals} Sky position, redshift, and flux densities for all members of the HCG and MLCG systems detected in the LoTSS.}\\\\
\hline
\hline
Name            &       Comp    &       RA(deg) &       DE(deg) &       z       &       Si6(mJy)        &       Si6err(mJy)     \\
\hline
\endfirsthead
\caption{continued.}\\
\hline\hline
Name            &       Comp    &       RA(deg) &       DE(deg) &       z       &       Si6(mJy)        &       Si6err(mJy)     \\
\hline
\endhead
\hline
\endfoot
HCG56           &       a       &       173.14723       &       52.94724        &       0.0268  &       ?       &       ?       \\
                        &       b       &       173.13655       &       52.93929        &       0.0268  &       ?       &       ?       \\
                        &       c       &       173.19458       &       52.94088        &       0.0275  &       ?       &       ?       \\
                        &       d       &       173.16774       &       52.9504         &       0.0265  &       ?       &       ?       \\
                        &       e       &       173.15305       &       52.94767        &       0.0271  &       ?       &       ?       \\
HCG60           &       a       &       180.77867       &       51.68549        &       0.0633  &       2.3     &       0.4     \\
                        &       b       &       180.77957       &       51.67521        &       0.0634  &2402.8 &  90.5   \\
                        &       c       &       180.7361        &       51.69548        &       -               &       3.6     &       0.2     \\
MLCG8           &       a       &       223.76506       &       54.32933        &       0.0996  &       0.9     &       0.1     \\
MLCG10          &       c       &       188.84366       &       47.70836        &       0.0287  &       5.9     &       0.3     \\
MLCG23          &       b       &       207.74294       &       56.04366        &       0.0681  &       2.5     &       0.2     \\
MLCG24          &       a       &       207.92557       &       55.99533        &       0.0691  &       *       &       *       \\
                        &       b       &       207.92549       &       56.00126        &       0.0703  &       *       &       *       \\
                        &       c       &       207.92726       &       55.99937        &       0.0678  & 925.7   &  46.3 \\
MLCG28          &       a       &       211.8439        &       54.00246        &       0.1162  &       ?       &       ?       \\
MLCG39          &       a       &       188.82079       &       47.29997        &       0.031   &       0.4     &       0.1     \\
                        &       c       &       188.84328       &       47.29448        &       0.0308  &       0.7     &       0.1     \\
MLCG41          &       a       &       199.9581        &       52.06146        &       0.0156  &       2.7     &       0.1     \\
                        &       b       &       199.94809       &       52.07033        &       0.0154  &       111     &       0.6     \\
                        &       c       &       200.00409       &       52.0509         &       0.0154  &       1.6     &       0.1     \\
MLCG44          &       a       &       229.85564       &       49.50135        &       0.0373  &       0.7     &       0.1     \\
MLCG45          &       a       &       198.9381        &       55.7888         &       0.0811  &       1.9     &       0.1     \\
                        &       b       &       198.93112       &       55.79943        &       0.0796  &       2.1     &       0.1     \\
                        &       c       &       198.932         &       55.78457        &       0.0808  &       2.6     &       0.1     \\
MLCG51          &       b       &       197.82474       &       47.5465         &       0.0289  &       4.9     &       0.3     \\
MLCG118         &       c       &       223.5346        &       46.9845         &       0.0873  &       ?       &       ?       \\
MLCG176         &       b       &       216.33788       &       47.2453         &       0.072   &       0.6     &       0.1     \\
                        &       d       &       216.33478       &       47.25852        &       0.075   &       7.5     &       0.4     \\
MLCG177         &       b       &       209.63158       &       49.53708        &       0.1053  &       0.4     &       0.1     \\
MLCG1108        &       a       &       171.63628       &       55.57245        &       0.0471  &       1.5     &       0.1     \\
MLCG1116        &       b       &       163.54654       &       54.83838        &       0.0732  &       ?       &       ?       \\
MLCG1117        &       a       &       163.40237       &       54.86794        &       0.0716  & 15.7    &       0.8     \\
MLCG1121        &       b       &       170.76057       &       56.02791        &       0.0561  &       1.7     &       0.1     \\
                        &       c       &       170.77632       &       56.01753        &       0.0593  &       ?       &       ?       \\
MLCG1130        &       a       &       178.69844       &       52.82193        &       0.0677  &       0.6     &       0.1     \\
MLCG1131        &       c       &       162.44643       &       51.8942         &       0.0247  &       ?       &       ?       \\
MLCG1138        &       a       &       166.99965       &       55.88076        &       0.0505  &       ?       &       ?       \\
MLCG1149        &       a       &       164.99811       &       50.05664        &       0.0236  &       0.8     &       0.1     \\
                        &       c       &       164.97108       &       50.01519        &       0.0257  &       ?       &       ?       \\
MLCG1167        &       c       &       177.89798       &       54.22895        &       0.0648  &       5.1     &       0.3     \\
MLCG1186        &       b       &       172.46236       &       50.14433        &       0.0469  &       0.9     &       0.1     \\
                        &       c       &       172.48874       &       50.13555        &       0.0478  &       1.5     &       0.1     \\
MLCG1191        &       a       &       177.64488       &       50.52897        &       0.023   &       1.2     &       0.1     \\
                        &       b       &       177.6297        &       50.54302        &       0.0239  &       1.4     &       0.1     \\
MLCG1192        &       a       &       197.03194       &       50.07972        &       0.055   &       0.8     &       0.1     \\
                        &       d       &       197.01701       &       50.06569        &       0.0554  &       3.3     &       0.2     \\
MLCG1193        &       a       &       199.12987       &       49.91392        &       0.0487  &       0.7     &       0.1     \\
                        &       b       &       199.11877       &       49.92035        &       0.048   &       ?       &       ?       \\
                        &       c       &       199.15527       &       49.91113        &       0.0482  &       0.6     &       0.1     \\
MLCG1195        &       a       &       217.10788       &       45.96946        &       0.0738  &       ?       &       ?       \\
                        &       b       &       217.09018       &       45.96586        &       0.0749  &       ?       &       ?       \\
MLCG1202        &       a       &       173.06662       &       55.96308        &       0.0515  &       0.5     &       0.1     \\
                        &       b       &       173.09668       &       55.96744        &       0.0515  &       *       &       *       \\
                        &       c       &       173.09479       &       55.97182        &       0.0501  & 512.8   & 25.6  \\
                        &       d       &       173.09923       &       55.96771        &       0.0511  &       *       &       *       \\
MLCG1204        &       a       &       178.0392        &       53.3167         &       0.0558  &       2.4     &       0.2     \\
                        &       b       &       178.02596       &       53.31119        &       0.0553  &       10.1&   0.5     \\
                        &       c       &       178.01091       &       53.31418        &       0.055   &       1.5     &       0.1     \\
MLCG1208        &       c       &       178.58125       &       52.69461        &       0.063   &       ?       &       ?       \\
MLCG1216        &       a       &       208.16902       &       48.7993         &       0.0683  &       0.7     &       0.1     \\
                        &       c       &       208.19493       &       48.8071         &       0.0696  &       0.2     &       0.1     \\
MLCG1218        &       a       &       187.72652       &       51.34129        &       0.0441  &       2.1     &       0.1     \\
                        &       b       &       187.74174       &       51.34415        &       0.0441  &       1.2     &       0.1     \\
                        &       c       &       187.71458       &       51.33944        &       0.0451  &       0.4     &       0.1     \\
                        &       d       &       187.75154       &       51.35918        &       0.0425  &       0.5     &       0.1     \\
MLCG1225        &       b       &       177.85884       &       56.4074         &       0.0184  &       3.7     &       0.2     \\
MLCG1227        &       b       &       177.02457       &       54.64628        &       0.0599  &       8.4     &       0.4     \\
MLCG1246        &       b       &       193.51028       &       48.78505        &       0.0378  &       9.2     &       0.5     \\
MLCG1258        &       c       &       214.86084       &       45.84233        &       0.0262  &       3.4     &       0.2     \\
MLCG1259        &       b       &       172.28679       &       54.11541        &       0.0681  &       0.6     &       0.1     \\
MLCG1262        &       a       &       169.28462       &       54.91743        &       0.1369  &       0.2     &       0.1     \\
                        &       b       &       169.27672       &       54.91716        &       0.1391  &       2.8     &       0.2     \\
MLCG1264        &       a       &       164.04887       &       46.88068        &       0.0286  &       0.3     &       0.1     \\
                        &       b       &       164.03908       &       46.89496        &       0.0291  &       5.4     &       0.3     \\
                        &       c       &       164.07512       &       46.86539        &       0.0285  &       1.8     &       0.1     \\
MLCG1266        &       b       &       169.86061       &       54.37024        &       0.072   &       0.3     &       0.1     \\
MLCG1267        &       a       &       224.93533       &       47.94507        &       0.0312  &       ?       &       ?       \\
MLCG1270        &       b       &       181.97426       &       51.60904        &       0.0885  &       1.9     &       0.1     \\
MLCG1276        &       c       &       184.66115       &       48.85408        &       0.0448  &       0.5     &       0.1     \\
MLCG1278        &       b       &       221.22517       &       48.64193        &       0.0501  &       2.1     &       0.1     \\
                        &       c       &       221.24231       &       48.64377        &       0.0496  &  22.7   &       0.1     \\
MLCG1283        &       a       &       186.95001       &       50.38782        &       0.04    &  12.6   &       0.6     \\
                        &       b       &       186.95023       &       50.40793        &       0.0388  &       1.9     &       0.1     \\
MLCG1284        &       a       &       185.42297       &       55.78717        &       0.0343  &       1.2     &       0.1     \\
MLCG1296        &       a       &       173.93216       &       49.16663        &       0.0353  &       1.1     &       0.1     \\
MLCG1299        &       c       &       177.927         &       54.94741        &       0.0611  &       ?       &       ?       \\
MLCG1309        &       a       &       179.18962       &       55.0303         &       0.0638  &       0.6     &       0.1     \\
                        &       c       &       179.22527       &       55.04219        &       0.064   &       1.1     &       0.1     \\
MLCG1310        &       a       &       166.61151       &       48.63585        &       0.0246  &       0.9     &       0.1     \\
                        &       b       &       166.63048       &       48.65162        &       0.0245  &  23.4   &       1.2     \\
MLCG1322        &       c       &       226.91977       &       52.83908        &       0.1103  &       1.1     &       0.1     \\
MLCG1323        &       a       &       219.6744        &       56.0012         &       0.0428  &       0.6     &       0.1     \\
                        &       b       &       219.66103       &       56.00018        &       0.043   &       0.6     &       0.1     \\
MLCG1324        &       b       &       220.31053       &       55.85105        &       0.0508  &       5.0     &       0.3     \\
MLCG1326        &       a       &       170.00888       &       47.2397         &       0.1097  &       ?       &       ?       \\
MLCG1332        &       d       &       168.76317       &       54.48249        &       0.067   &       8.5     &       0.4     \\
MLCG1334        &       a       &       171.6179        &       54.78723        &       0.0471  &       67.5&   3.4     \\
                        &       b       &       171.60854       &       54.79143        &       0.0463  &       1.5     &       0.1     \\
                        &       d       &       171.63957       &       54.77358        &       0.0469  &       0.5     &       0.1     \\
MLCG1335        &       a       &       173.90157       &       54.94856        &       0.0193  &       0.3     &       0.1     \\
MLCG1349        &       a       &       185.15677       &       47.64976        &       0.0588  &       7.7     &       0.4     \\
                        &       c       &       185.18103       &       47.66221        &       0.0606  &       0.9     &       0.1     \\
MLCG1350        &       c       &       225.30052       &       47.27317        &       0.0873  &       8.0     &       0.5     \\
MLCG1361        &       c       &       174.47816       &       47.46637        &       0.0341  &       1.5     &       0.1     \\
MLCG1362        &       c       &       170.96149       &       47.18843        &       0.0536  &       4.5     &       0.2     \\
MLCG1364        &       a       &       205.93755       &       55.63378        &       0.0682  &       3.0     &       0.2     \\
                        &       c       &       205.95491       &       55.62398        &       0.0673  &       0.5     &       0.1     \\
                        &       d       &       205.91359       &       55.63152        &       0.0689  &       0.7     &       0.1     \\
MLCG1365        &       a       &       205.80719       &       55.62277        &       0.0663  &       1.7     &       0.1     \\
                        &       c       &       205.8091        &       55.60314        &       0.0679  &       17.4&   0.9     \\
MLCG1370        &       a       &       205.91328       &       55.37853        &       0.0676  &       1.0     &       0.1     \\
MLCG1371        &       a       &       181.22063       &       54.23176        &       0.0503  &       5.3     &       0.3     \\
MLCG1372        &       a       &       213.35889       &       54.73987        &       0.0834  &       4.0     &       0.2     \\
                        &       b       &       213.36392       &       54.72941        &       0.0828  &       0.5     &       0.1     \\
                        &       c       &       213.35883       &       54.72194        &       0.0819  &       1.2     &       0.1     \\
MLCG1373        &       c       &       221.22308       &       51.34109        &       0.089   &       0.3     &       0.1     \\
                        &       d       &       221.21843       &       51.33265        &       0.0888  &       0.8     &       0.1     \\
MLCG1374        &       a       &       222.5684        &       50.79213        &       0.0702  &       *       &       *       \\
                        &       b       &       222.57103       &       50.78824        &       0.0696  &       25.2&   1.2     \\
                        &       c       &       222.571         &       50.78339        &       0.0689  &       *       &       *       \\
                        &       d       &       222.55338       &       50.77462        &       0.0702  &       1.7     &       0.1     \\
MLCG1375        &       a       &       204.5809        &       56.40196        &       0.0523  &       2.4     &       0.1     \\
MLCG1379        &       a       &       218.30403       &       52.96314        &       0.0473  &       8.9     &       0.5     \\
                        &       c       &       218.29398       &       52.97516        &       0.0476  &       2.0     &       0.1     \\
MLCG1380        &       a       &       178.97305       &       48.28494        &       0.0318  &       2.2     &       0.1     \\
                        &       b       &       178.96626       &       48.25454        &       0.0323  &       0.7     &       0.1     \\
MLCG1495        &       b       &       219.7267        &       46.66561        &       0.0359  &       3.0     &       0.2     \\
MLCG1562        &       a       &       174.78572       &       55.66449        &       0.0612  & 1358.2&  67.9   \\
                        &       b       &       174.80092       &       55.66598        &       0.0623  &       *       &       *       \\
MLCG1563        &       b       &       203.9413        &       47.42017        &       0.0618  &       14.2&   0.7     \\
MLCG1564        &       a       &       176.8391        &       55.73018        &       0.0515  &       3.1     &       0.2     \\
                        &       c       &       176.81519       &       55.71708        &       0.0519  &       2.2     &       0.1     \\
MLCG1565        &       a       &       176.90718       &       55.71792        &       0.0491  &       2.3     &       0.2     \\
MLCG1567        &       a       &       173.7054        &       49.07764        &       0.0333  &       12.9&   6.4     \\
MLCG1568        &       a       &       162.9318        &       51.02216        &       0.025   &       1.0     &       0.1     \\
                        &       c       &       162.97379       &       51.00651        &       0.0276  &       19.0&   1.0     \\
\hline
\end{longtable}
\tablefoot{
Column descriptions:\\
Name: name of the system;\\
Comp: companion designation, according to HyperLeda \citep{hyperleda};\\
RA, Dec: sky position (in degrees) of the system;\\
z: spectroscopic redshift from \citet{mlcg};\\
Si6: integrated (measured by \textsc{blobcat}) flux of the group;\\
        ?: flux unknown due to the presence of artefacts/low signal;\\
    *: flux unknown due to the affiliation with a larger radio structure;\\
Si6err: uncertainty of the latter.
}

\twocolumn

\begin{table*}[ht]
\caption{\label{tab:fluxes} Basic data on the subsample of galaxy groups.} 
\centering
\begin{tabular}{lccccccccc}
\hline
\hline
System          & $\rm N_G$     &       $z$             &               $D$             &       $\sigma_{V}$            &               $\Phi$  &       LoTSS-6                 &       LoTTS-20                &       FIRST                   &       NVSS            \\
  &  &  & [Mpc] & [km/s] & [kpc] & [mJy] & [mJy] & [mJy] & [mJy] \\
\hline
HCG 60          &       3               &       0.0625  &       255     &       421 $\pm$ 72                        &               38.6 $\pm$ 4    &       2409 $\pm$ 91         &       2444 $\pm$ 122   &     324 $\pm$ 14    &       655 $\pm$ 33        \\
MLCG 24         &       3               &       0.0691  &       282     &       353 $\pm$ 60                        &               22 $\pm$ 4.6    &       926 $\pm$ 46        &        934 $\pm$ 47    &      180 $\pm$ 9             &       223 $\pm$ 11        \\
MLCG 41         &       4               &       0.0154  &       63              &       33      $\pm$ 6                               &               36.8 $\pm$ 4    &       15 $\pm$ 1    &   14.3 $\pm$ 0.7   &       3.1$\pm$ 0.2*   &       4.7 $\pm$ 0.5     \\
MLCG 1374       &       3               &       0.0697  &       285             &       177 $\pm$ 36                        &               95.6 $\pm$ 15.8 &       27 $\pm$ 1    &   26.6 $\pm$ 1.3   &       4.8 $\pm$ 0.3   &       5.5 $\pm$ 0.5     \\

\hline
\end{tabular}
\tablefoot{
Description of columns:\\
$\rm N_{G}$ -- number of galaxies in the group;\\
$z$ -- redshift (spectroscopic);\\
$D$ -- approximate distance to the system, calculated from redshift, assuming $\rm H_0$ = 73 km/s/Mpc;\\
$\sigma_{V}$ -- velocity dispersion;\\
$\Phi$ -- physical diameter of the system (from optical data);\\
The last four columns contain the flux densities from the respective survey data.\\
* signal-to-noise ratio does not exceed 2.8.\\
}
\end{table*}

\section{Discussion}
\label{sec_discussion}

\subsection{Magnetic field strength}

The strength of the total magnetic field contained inside each of the sample systems was estimated from LoTSS 150\,MHz data, using the formulae presented by \citet{bfeld} and assuming the revised equipartition of energy between the cosmic-ray electrons and the magnetic field principle, which is believed to be kept even in case of starburst activity \citep{lacki13}. The proton-to-electron ratio is fixed to the typical value of 100; however, the strength and energy density are weakly dependent on this parameter. The final estimates include twice the uncertainties of the flux density at 150\,MHz and the spectral index measurements. Pathlengths (d) are defined individually for each of the cases (and discussed in the following paragraphs). To calculate the strength of the magnetic field in each of the systems, we used the flux densities from Table~\ref{tab:fluxes}, assuming that the 150\,MHz flux density is purely non-thermal, and a constant, 10\% thermal fraction at 1400\,MHz, which, according to \citealt{niklas97}, is the typical value for spiral galaxies. These values were transformed into intensities, taking into account the extent of the radio emission.\\

Unfortunately, it is not possible to calculate the spectral index distribution. The snapshot character of both the FIRST and NVSS results in a much sparser sampling of the $(u,v)$ plane in these data. In addition, the loose B-configuration of the Very Large Array interferometer used in the former survey introduces missing zero-spacing flux problem. Owing to this incompatibility, it is only possible to either derive the spectral index for point sources (using FIRST data) or its integrated value for the whole systems (using NVSS). As an analysis of the point sources, such as the active cores of MLCG\,24 and HCG\,60, is beyond the scope of this paper, we only derive the integrated, non-thermal spectral indices. Parameters used in the calculation of the magnetic field, as well as the estimated strengths, are placed in Table~\ref{tab:magfield}.\\

\begin{table*}[ht]
\caption{\label{tab:magfield}Parameters used in magnetic field estimation and derived strength.}
\begin{tabular}{lccccc}
\hline
\hline
System          &       $\rm S_{150}$ [mJy]     &       $\alpha_{NTH}$  &       d [kpc]           &       $\rm B_{TOT}$ [$\mu$G]  & $\tau$ [Gyr]          \\
\hline
H60             &       2425 $\pm$ 244                  &       0.64 $\pm$ 0.07    &       55 -- 110       &       6.8 $\pm$ 1.1                   & 0.22 $\pm$ 0.05         \\
M24             &        930 $\pm$ 94                   &       0.63 $\pm$ 0.07    &       20 --30         &       9.0 $\pm$ 1.0                   & 0.14 $\pm$ 0.03         \\
M41             &       14.8 $\pm$ 1.4                  &       0.56 $\pm$ 0.09    &       4 -- 20         &       8.0 $\pm$ 3.0                   & 0.22 $\pm$ 0.11         \\
M1374   &       26.8 $\pm$ 2.6                  &       0.76 $\pm$ 0.09 &       4 -- 25           &       5.9 $\pm$ 1.6                   & 0.30 $\pm$ 0.12               \\

\hline
\end{tabular}
\tablefoot{
Description of columns:\\
$\rm S_{150}$: total flux density at 150\,MHz, averaged value of the 6 and 20 arcsec maps;\\
$\alpha_{NTH}$: non-thermal spectral index calculated between 150 and 1400\,MHz;\\
d: pathlength through the medium that hosts the magnetic field;\\
$\rm B_{TOT}$: magnetic field strength;\\
$\tau$ synchrotron age.
}
\end{table*}

The magnetic field strength for HCG\,60 was estimated from the 1400 and 4860\,MHz data by \citet{bnw17} as $3.2 \pm 1.7$\,$\mu$G. Owing to the fact that this study used mainly low-resolution, high-frequency data to overcome the possible missing zero-spacings flux issue, the pathlength through the source was estimated as 100--300\,kpc, assuming spherical symmetry. Analysis of high-resolution LoTSS data (Fig.~\ref{fig:hcg60}) suggests that the radio envelope contains two separate lobes of the radio galaxy; therefore, the assumed pathlength should be at most 110\,kpc. Taking half of this value as the lowest estimate of this parameter yields a total magnetic field strength of $\rm 6.8 \pm 1.1$\,$\mu$G. This accounts for twice the previously known value \citep{bnw17}. However, in the previous study it was possible to disentangle the emission from extended and compact sources, this being a probable reason for the observed discrepancy. For MLCG\,24 a similar approach was chosen. Under the assumption of cylindrical symmetry and using the width of the radio jets as an estimate of the diameter, the pathlength is fixed as 20--30\,kpc. Here the estimated strength of the magnetic field is $\rm 9.0 \pm 1.0$\,$\mu$G. Given the large size of the radio structures, the minimum velocities of the electrons necessary so they would be transferred from the regions of acceleration (cores) to the outskirts of the IGM are approximately 500\,km/s for MLCG\,24 and 650\,km/s for HCG\,60; this is comparable with the estimates given for HCG\,15 by \citet{bnw17}.\\

For MLCG\,41 and MLCG\,1374 a common method of the pathlength estimation was also used. Similar to AGN-hosting groups, the cylindrical symmetry is used -- in this case as the upper limit (20\,kpc for the former and 25\,kpc for the latter). The lower value is set to the expected scaleheight of the radio-emitting disc of the face-on galaxies present in both of these groups. This parameter can be different for different systems: \citet{heesen09} gave the value of 1.7\,kpc for the thick disc of NGC\,253, \citet{beck91} argued that in case of NGC\,4631 the value is equal to 1.9\,kpc (later assumed to be app. 2.3\,kpc; see \citealt{krause14}), and for NGC\,891 the scaleheight should be approximately 4\,kpc. Finally, \citet{soida11} estimated this parameter to be as large as 8\,kpc in case of NGC\,5575. We decided that in both cases the lower estimate is fixed as 4\,kpc, so it moreover falls between the maximal and minimal reported values. The strengths of the total magnetic field for these systems are then $\rm 5.9 \pm 1.6$\,$\mu$G for MLCG\,1374 and $\rm 8 \pm 3$\,$\mu$G for MLCG\,41. Minimum velocities of electrons are of an order of magnitude lower than in case of HCG\,60 and MLCG\,24: 60\,km/s for MLCG\,41 and 100\,km/s for MLCG\,1364. However, these values describe the mixed input from galaxies and the space between them; it is desirable to derive values for the intergalactic space only. Whereas flux density estimates for the extended filaments detected in the LoTSS have already been presented in Sections~\ref{results_mlcg41} and \ref{results_mlcg1374}, lower sensitivity at 1400\,MHz makes it difficult to provide the information necessary to calculate the spectral index of these features. In order to overcome these hindrances and provide an, albeit rough, estimate for the intergalactic magnetic field in both of these systems, we made an assumption that the spectrum of the emission associated with the IGM is rather steep; it has an index of app. 1. Other parameters were left unchanged. Re-doing the estimates with these new inputs yields total magnetic fields as strong as $\rm 4.9 \pm 1.1$\,$\mu$G for MLCG\,1374 and $\rm 5.8 \pm 1.3$\,$\mu$G for MLCG\,41. Albeit somewhat lower than derived before, these are still considerable strengths, meaning that the observed radio envelopes could easily form by transporting electrons from actively star-forming galaxies.\\

In all of the cases, the derived mean magnetic field strength is generally comparable to what is expected to be found in field galaxies, as \citet{niklas95} calculates the typical value for spiral galaxies to be equal to $9 \pm 1.3$\,$\mu$G. These values are somewhat lower at most half of an order of magnitude than that found inside the intergalactic regions of the Stephan's Quintet \citep{xu03, bnw13B} or inside the radio bridges of Taffy I and Taffy II systems \citep{drzazga11}. In all the above literature examples, the magnetic field is expected to be able to alter the gas dynamics due to the input of the energy contained within. Owing to that fact, we can then conclude that a similar situation is likely to happen in the four sample systems from this study. It should be noted that this energy output is unlikely to disturb the global movement of the gas (due to the rotation of the galaxy), but it certainly influences its local behaviour.

\subsection{How widespread is the radio emission in galaxy groups?}

As mentioned in Sect.~\ref{sec_intro}, not much is known about the radio emission in galaxy groups. The only survey study of a large set of galaxy groups -- that of \citet{menon85} -- revealed only one extended radio structure in a galaxy group, and 32 additional systems in which galaxy-bound emission was found (in 44 galaxies). This would mean that only 33\% of galaxy groups are sources of radio emission and only 10\% of galaxies in groups are detectable in the radio regime. Compared to these numbers, LoTSS data seem to reveal a lot more detections: approximately two-thirds of galaxy groups from our sample are radio emitters, and around 30\% of all galaxies are detected. Furthermore, as many of 17 galaxy groups exhibit extended radio structures at 150\,MHz, compared to just one detected by \citet{menon85} at 1635\,MHz, meaning that this type of radio emission is not as unique, as originally thought; this conclusion is supported by the previous findings (e.g. by \citealt{giacintucci11}, \citealt{bnw13B}, or \citealt{bnw17}). A direct comparison of the results can be done for HCG\,56 and HCG\,60, which are included in both of the studies. While at 1635\,MHz each of these galaxy groups hosts just a single radio-emitting galaxy, all members of HCG\,60 and 4 (out of 5) members of HCG\,56 are detected at 150\,MHz. There are also hints of an extended structure in HCG\,56, while it was just a point source at 1635\,MHz.\\

It is also worthwhile to compare the detection rate to that estimated by \citet{paul18}. Depending on the given detection limit, we can expect to detect radio emission in 3--21\% of the studied systems; the exact flux density depends on a number of parameters, for example the mass of the host system. The success ratio of 14\% acquired with LoTSS data fits well with the predictions; however, to fully evaluate the mass-flux density relation given by \citet{paul18}, a more detailed study is needed. This will be included in a subsequent paper.

\section{Conclusions}
\label{sec_conclusions}
We have carried out a pilot study of the radio emission from galaxy groups using LoTSS data. As many as 73 galaxy groups out of 120 are detected in the radio continuum at 150\,MHz (possibly 7 more). Out of these samples, 17 host intergalactic, radio-emitting structures, signifying the presence of intergalactic magnetic fields in galaxy groups. Eight additional systems are probable emitters. Out of 419, 110  galaxies are found as certain, radio-emitting galaxies; an additional 26 stand for possible detections as well. This means that an average number density of radio-emitting galaxies is 0.92--1.13 per group. All these numbers are significantly higher than those presented by and calculated from the data of \citet{menon85}, suggesting that continuum radio emission (especially the extended one) is much more widespread, than previously thought. A detailed study of four objects (selected among the aforementioned 17) suggests that these systems -- HCG\,60, MLCG\,24, MLCG\,41, and MLCG\,1374 -- host relatively strong ($\approx 6 - 9$\,$\mu$G) magnetic fields. This strength is comparable both with typical values reported for spiral galaxies by \citet{niklas95}, and those of a small sample of galaxy groups studied by \citet{bnw17}. At the same time, they are somewhat lower that those found for intergalactic structures inside Taffy systems (interacting pairs with member galaxies connected by a radio continuum bridge, \citealt{drzazga11}). As the physical conditions in the gas, for example a lack of strong thermal sources, inside MLCG\,41 and MLCG\,1374, are likely to be similar to that of the Taffy systems, it can be expected that the magnetic fields have an impact on the local gas dynamics of these systems. Compared to the higher frequency observations, LoTSS data for these four systems show similar or larger extent in the radio continuum, while the accuracy in portraying the small-scale structures is on a comparable, or higher level, providing a unique chance to disentangle particular radio emitters from their surroundings.

\begin{acknowledgements}
We would like to thank the anonymous referee for suggestions that helped in preparing the final version of this paper. Also, we are indebted to Olaf Wucknitz from the MPIfR Bonn for providing us with useful comments that helped to improve this paper. This work has received support from the Polish National Science Centre (NCN), grant no. UMO-2016/23/D/ST9/00386. The work at Ruhr-University Bochum is supported by BMBF Verbundforschung under D-LOFAR IV - FKZ: 05A17PC1. The data used in this work was in part processed on the Dutch national e-infrastructure with the support of SURF Cooperative through grant e-infra 160022 \& 160152. LOFAR, the LOw Frequency ARray designed and constructed by ASTRON, has facilities in several countries, which are owned by various parties (each with their own funding sources) and are collectively operated by the International LOFAR Telescope (ILT) foundation under a joint scientific policy. The project has in part also benefited from the exchange programme between Jagiellonian University in Krak\'ow and Ruhr-University Bochum.  
\end{acknowledgements}


\end{document}